\documentclass[iop,apj]{emulateapj}
\usepackage{natbib}
\usepackage{amsmath}
\usepackage{url}
\slugcomment{Published in ApJ, 790, 93; 2014 Aug 1}
\shorttitle{Wideband Timing of Radio Pulsars}
\shortauthors{Pennucci, Demorest \& Ransom}

\begin{document}

\title{Elementary Wideband Timing of Radio Pulsars}

\author{Timothy T. Pennucci\footnote{TTP is a graduate student at the National Radio Astronomy Observatory}}
\affil{University of Virginia, Department of Astronomy, PO Box 400325 Charlottesville, VA 22904-4325\\TTP is a graduate student at the National Radio Astronomy Observatory}
\email{tim.pennucci@nanograv.org}
\and
\author{Paul B. Demorest \& Scott M. Ransom}
\affil{National Radio Astronomy Observatory, 520 Edgemont Road, Charlottesville, VA 22903-2475}
\email{pdemores@nrao.edu, sransom@nrao.edu}

\begin{abstract}
We present an algorithm for the simultaneous measurement of a pulse time-of-arrival (TOA) and dispersion measure (DM) from folded wideband pulsar data.  We extend the prescription from Taylor (1992) to accommodate a general two-dimensional template ``portrait'', the alignment of which can be used to measure a pulse phase and DM.  We show that there is a dedispersion reference frequency that removes the covariance between these two quantities, and note that the recovered pulse profile scaling amplitudes can provide useful information.  We experiment with pulse modeling by using a Gaussian-component scheme that allows for independent component evolution with frequency, a ``fiducial component'', and the inclusion of scattering.  We showcase the algorithm using our publicly available code on three years of wideband data from the bright millisecond pulsar J1824-2452A (M28A) from the Green Bank Telescope, and a suite of Monte Carlo analyses validates the algorithm.  By using a simple model portrait of M28A we obtain DM trends comparable to those measured by more standard methods, with improved TOA and DM precisions by factors of a few.  Measurements from our algorithm will yield precisions at least as good as those from traditional techniques, but is prone to fewer systematic effects and is without ad hoc parameters.  A broad application of this new method for dispersion measure tracking with modern large-bandwidth observing systems should improve the timing residuals for pulsar timing array experiments, like the North American Nanohertz Observatory for Gravitational Waves.
\end{abstract}
%\keywords{B1821-24, J1824-2452A, M28A, millisecond pulsars, pulsar timing, wideband timing}

%\notetoeditor{}

\section{Introduction}
\label{intro}

The practice of pulsar timing attempts to model the rotation of a neutron star by phase-connecting periodic observations of its pulsed, broadband radio signal.  The earliest demonstration of long-term timing observations came relatively soon after the discovery of pulsars~\citep{Roberts&Richards71, Hewish68}.  The scientific merits garnered from pulsar timing span astrophysical fields such as planetary science, the interstellar medium, nuclear physics, gravitational wave physics, and are all well-documented (for a review, see eg. Chapter 2,~\citet{L&K04}).

Pulsar timing and its related experiments have carved out a ``sweet spot'' in the radio frequency regime that naturally emerges as a trade-off between the pulsars' steep power-law spectra at the high-frequency end, and the low-frequency drawbacks arising from the pulsed radio signal having to propagate through the ionized interstellar medium (ISM) and the Earth's ionosphere, as well as having to compete with the diffuse background of the galactic synchrotron continuum.  The latter has a spectral index in the 1--10~GHz range of $\approx-2.8$~\citep{Platania98}.  Population studies have shown that pulsars have an average spectral index around -1.4 at gigahertz frequencies~\citep{Bates13}.  The most relevant ISM effect arises from propagation through a homogeneously ionized medium.  Interstellar dispersion alters the group-velocity of the radio signal, retarding the arrival of pulses emitted at a radio frequency $\nu$ by a time $t_{\textrm{DM}}$ (relative to an infinite frequency signal) according to the cold-plasma dispersion law,
\begin{equation}
\label{DMlaw}
    t_{\textrm{DM}} = K \times \textrm{DM} \times \nu^{-2},
\end{equation}
where $K \equiv \frac{e^2}{2\pi m_ec} = 4.148808(3) \times 10^{3}$~MHz$^2$~cm$^{3}$~pc$^{-1}$~sec is called the dispersion constant\footnote{$K$ is a combination of the electron charge $e$, electron mass $m_e$, and speed of light $c$.  It is common practice in the pulsar community to adopt the approximation $K^{-1} = 2.41\times10^{-4} $~MHz$^{-2}$~cm$^{-3}$~pc~sec$^{-1}$~\citep{L&K04}, which we have used in~\S\ref{m28a} and~\S\ref{mcs}.}, and DM is the dispersion measure.  The dispersion measure is defined as
\begin{equation}
\label{DM}
    \textrm{DM} \equiv \int\limits_l n_e\,dl,
\end{equation}
which is the free-electron column density along the path-of-propagation $l$ to the pulsar.  The pulse-broadening effect of multi-path propagation through a turbulent, inhomogeneous ISM, known as interstellar scattering, has an even stronger spectral index $\approx-4$, and becomes increasingly important at lower frequencies for the highest-DM, farthest pulsars~\citep{L&K04}.  Scattering not only broadens the pulsed signal, but delays an intrinsically sharp pulse by an amount roughly proportional to its width, and so is a source of bias in timing measurements.  The determination of dispersion measures and effects from scattering have been non-trivial problems concomitant with timing measurements since the beginning~\citep{Rankin70, Rankin71}.

Nearly all observations taken for (high-precision) pulsar timing experiments are taken within the radio window mentioned above, which lies somewhere in the two decades bounded by about 100~MHz and 10~GHz.  The middle decade centered around 1500~MHz seems to be the perennial favorite for timing experiments.  Recent developments in pulsar instrumentation and computing over the last 5--10 years have enabled more accurate and sensitive timing measurements.  Namely, coherent dedispersion, which completely removes the quadratic time-delay due to a known amount of interstellar dispersion~\citep{Hank&Rick75}, required significant advances in computer technology before becoming feasible in real-time on a wide-bandwidth signal.

Historically, observations that implemented coherent dedispersion were limited by computing resources to a bandwidth of order $\sim$100~MHz or less, which is less than most receiver bandwidths.  Thus, if one wanted to cover a large portion of the pulsar spectrum, either for timing, spectral, or interstellar medium purposes, several adjacent receiver bands had to be observed separately, which often meant asynchronous measurements and non-contiguous frequency coverage.   The implementation of real-time coherent dedispersion to large, instantaneously observed bandwidths has led to the regime wherein the receiver bandwidth (BW) is a limiting factor.  The first generation of GHz-bandwidth, coherent dedispersion instruments has been proliferating in the pulsar community for the past several years, beginning with the Green Bank Ultimate Pulsar Processing Instrument (GUPPI)\footnote{\url{www.safe.nrao.edu/wiki/bin/view/CICADA/NGNPP}} outfitted for the 100-m Robert C. Byrd Green Bank Telescope (GBT)\footnote{The National Radio Astronomy Observatory is a facility of the National Science Foundation operated under cooperative agreement by Associated Universities, Inc.}~\citep{Duplain08}.  GUPPI is an FPGA- and GPU-based system capable of real-time coherent dedispersion of an 800~MHz bandwidth.

The smearing $\delta t_{\textrm{DM}}$ incurred from incorrectly dedispersing a narrow frequency-channel of bandwidth $\Delta\nu = \frac{\textrm{BW}}{n_{chan}}$ and center frequency $\nu_c$ by an amount $\delta{\textrm{DM}}$ goes as
\begin{equation}
\begin{split}
\label{DMsmear}
    &\delta t_{\textrm{DM}} \approx \frac{2K\ \delta{\textrm{DM}}\ \Delta\nu}{\nu_c^3} \\
    &\qquad \approx 8.3\ \Big(\frac{\delta{\textrm{DM}}}{\textrm{cm}^{-3}\ \textrm{pc}}\Big)\Big(\frac{\Delta\nu}{1\ \textrm{MHz}}\Big)\Big(\frac{\nu_c}{1\ \textrm{GHz}}\Big)^{-3}\ \mu\textrm{s}.
\end{split}
\end{equation}
This equation demonstrates why it was difficult to obtain precise, broadband measurements from millisecond pulsars (MSPs) prior to coherent dedispersion.  Incoherent dedispersion shifts individual frequency channels of the data based on the assumed DM without compensating for dispersion within each channel.  This means that $\delta{\textrm{DM}}$ is equivalent to the full, true DM and so the dispersive smearing $\delta t_{\textrm{DM}}$ can easily be a significant fraction of the pulsar's spin period ($P_{s} \lesssim 10$~ms for MSPs).  To mitigate this problem, a large number of filterbank channels is needed, but it comes at the expense of a degraded time resolution due to Nyquist sampling constraints.  In turn, poorer time resolution means a poorer timing precision.

Moreover, tracking the dispersion measure changes in MSP observations is necessary for minimizing the timing residuals used in gravitational wave searches with a pulsar timing array (PTA)~\citep{You07}.  Equation~\ref{DMlaw} says that an incorrect DM of $10^{-3}$~cm$^{-3}$~pc at 1500~MHz introduces a delay of $\sim2 \mu$s relative to infinite frequency, which is well above the timing quality desired by PTA experiments.  As part of the Parkes Pulsar Timing Array\footnote{\url{www.atnf.csiro.au/research/pulsar/array/}} project~\citep{Hobbs13}, \citet{Keith13} developed a method to correct for inaccurate dispersion measures based on modeling the multi-frequency timing residuals.  However, the authors also postulate that more accurate DM variations could be measured from wideband receivers, which ameliorate the difficulties of aligning pulsar data taken with different receivers in different epochs.

The desire for very broadband pulsar observations (i.e. with significantly high fractional bandwidths, $\gtrsim$1) necessitates new, unique receiver designs that can cover much of the frequency range once concatenated from disjoint observations.  Wideband receivers and their complimentary, real-time coherent dedispersion backends will quickly facilitate developments in all realms of pulsar astrophysics, including studies of the pulsar spectrum, magnetosphere, and ISM properties.  One such instrument, called the Ultra-Broad-Band (UBB) receiver and associated backend\footnote{\url{www3.mpifr-bonn.mpg.de/staff/pfreire/BEACON.html}}, has been recently installed at the Effelsberg 100-m Telescope and covers a frequency range from $\sim$600~--~$\sim$3000~MHz.  

However, the current method for making pulse time-of-arrival (TOA) measurements that is used almost ubiquitously in the pulsar timing community does not use all of the information contained in new broadband observations.  In summary, the protocol employs frequency-averaged pulse profiles as models of the pulsar's signal for entire receiver bands, which ignores any profile evolution intrinsic to the pulsar or imposed by the ISM.  Both intrinsic profile evolution and DM changes are usually taken into account in the timing model for the pulsar's rotation, but there is no modeling of the effects from scattering or scintillation.  Arbitrary phase-offsets (known as ``JUMPs'') are introduced to align disparate template profiles that are used to measure TOAs from different frequency bands.  Multi-channel TOAs are also parameterized by both a quadratic delay (proportional to the DM) and an arbitrary function to remove residual frequency structure from otherwise unmodeled profile evolution.  Additionally, many multi-channel TOAs that are adversely affected by scintillation are often cut from being included in the timing model fit due to their inaccurately determined uncertainties; in effect, one is throwing away portions of the band that do contain much signal.

These methods are ad hoc and incomplete in that they were developed as the availability of bandwidth and multi-frequency observations became a ``problem'' (cf. ``the wide-bandwidth problem''~\citep{L&D13}), and were appropriate when observations covered a narrow bandwidth: phase JUMPs account for profile evolution occurring in frequency gaps that are not observed.  It seems natural in the era of wideband receivers --- when frequency evolution {\it is} observed in the band --- to devise a method for TOA measurement that includes a frequency-dependent model of the average pulse profile.  In doing so, it becomes straightforward to include a simultaneous measurement of the dispersion measure.  As we will show, a very simple extension to the algorithm that is currently used is a first step in a more comprehensive and necessary description of the received pulsar signal.

\section{The Algorithm}
\label{algo}

\subsection{Background}
\label{bkgd}

We assume that the recorded pulsar signal is cyclostationary for a given frequency, meaning the observed time-series data can be coherently folded modulo a pre-existing timing model to obtain an average signal shape that is stable with time.  This time-integrated light curve is often called a ``pulse profile'', which we label as $D(\varphi)$.  The quantity $\varphi = \varphi(t_{obs})$ represents the rotational phase of the neutron star at a particular moment in time $t_{obs}$, which is recorded by an observatory clock and later transformed into a more useful temporal coordinate system.

The central step in determining a pulse time-of-arrival is to measure the relative phase shift $\phi \in$ [-0.5, 0.5) between $D(\varphi)$ and a standard template profile, $P(\varphi)$, which is supposed to represent the noise-free average of the intrinsic pulse profile shape at the observed frequency.  In practice, the signal has been discretely and evenly sampled so that $D(\varphi)$ becomes $D(\varphi_j = (j+0.5)/n_{bin}) \equiv D_j$, where $j$ runs from 0 to $n_{bin}\!-\!1$, and $n_{bin}$ is the number of phase bins in the profile.  For pulsar timing purposes, the sampling time (and therefore the number of bins in the profile) is chosen to be appropriately small so that all meaningful information about the pulse profile with respect to the noise level is preserved in $D_j$.

The most obvious way to obtain a lag between $D_j$ and a template profile $P_j$ is to interpolate a maximum point in the discrete time-domain cross-correlation of the two functions.  However, Appendix A of~\citet{Taylor92} prescribes a Fourier frequency-domain technique for measuring the phase shift that has been used virtually ubiquitously for the past two decades in the pulsar timing community.  Besides the computational simplicity that is a consequence of the Fourier cross-correlation theorem, the reason for this ubiquity is because frequency domain techniques give very precise, accurate shifts for low duty-cycle pulsars, with uncertainties corresponding to less than a single time bin~\citep{Taylor92, Hotan05}.  Colloquially, this routine came to be known as FFTFIT, which is the designation we will use hereafter.  The advantage of FFTFIT is that a finite number of continuously-valued Fourier phases (instead of discrete time lags) are combined to interpolate a precise phase measurement.  An alternate formulation of FFTFIT can be found in Chapter 2 of~\citet{Demorest07}, which also recognizes that FFTFIT amounts to a cross-correlation completed in the frequency domain.  We have drawn from~\citet{Demorest07} as a starting point for the mathematical framework, and have borrowed some of its notation in what follows.

\subsection{Description}
\label{desc}

Because we are concerned with measurements of a wideband pulsar signal, we describe the observed pulse profile also as a function of frequency $\nu$, which we denote by $D(\nu, \varphi)$, and refer to as a ``pulse portrait''.  Similarly, the template portrait is $P(\nu, \varphi)$ and a simple model for the observed data is
\begin{equation}
\label{D}
    D(\nu, \varphi) = B(\nu) + a(\nu)P(\nu, \varphi-\phi(\nu)) + N(\nu, \varphi),
\end{equation}
where $\phi(\nu)$ will contain information about chromatic and achromatic phase shifts, $B(\nu)$ is effectively the bandpass shape of the receiver (analogously, $B$ is the ``DC''  or ``bias'' term when considering only a single frequency, as in FFTFIT), $a(\nu)$ is a multiplicative scale factor that can represent scintillation, and $N(\nu, \varphi)$ is additive noise.  $N(\nu, \varphi)$ is often assumed to be stationary and normally distributed with variance $\sigma^2(\nu)$, so that $N(\nu) \sim \textrm{Normal}(0, \sigma^2(\nu))$.  In the absence of radio-frequency interference (RFI), the noise in most pulse profiles is radiometer-noise dominated, which is highly Gaussian.  There are numerous methods for the removal of the bandpass shape $B(\nu)$ (which can be thought of as the frequency-dependent mean of the noise term $N(\nu)$), or one could follow an analogous treatment of the bias term in~\citet{Taylor92}.  One simple solution is to start all of the Fourier phase sums in the below equations at $k=1$, as we have done for our implementation.

Again, in practice the signal is discretized into $n_{bin}$ phase bins, but also into $n_{chan}$ frequency channels with center frequencies $\nu_n$.  We index each of the above frequency-dependent quantities with the letter $n$ (eg. $\phi_n, D_{nj}, P_{nj}$).  The question of determining $n_{chan}$ will be revisited in~\S\ref{mcs}.  The Discrete Fourier Transform (DFT) is a linear transformation, so taking a one-dimensional DFT of Equation~\ref{D} with respect to rotational phase $\varphi$, and making use of the discrete Fourier shift-theorem~\citep{Bracewell00} implies
\begin{equation}
\label{dnk}
    d_{nk} = a_np_{nk}e^{-2\pi ik\phi_n} + n_{nk},
\end{equation}
where  $i = \sqrt{-1}$, $k$ indexes the Fourier frequencies, and the DFT of a series $F_j$ is
\begin{equation}
\label{fk}
    f_k = \sum_{j=0}^{n_{bin}-1} F_je^{-2\pi ijk/n_{bin}}.
\end{equation}

The primary quantities of interest $\phi_n$, and the scaling parameters $a_n$ in Equation~\ref{dnk} can be found by minimizing the sum of the squares of the residuals between the data $d_{nk}$ and the shifted, scaled template $p_{nk}$, weighted by the noise\footnote{Assuming Gaussian noise, the noise variance $\sigma'^2_n$ in each frequency channel of $d_{nk}$ is greater than $\sigma^2_n$ in $D_{nj}$ by the factor $n_{bin}/2$.} in each frequency channel $\sigma'^2_n$.  In other words, we seek to minimize the statistic
\begin{equation}
\label{chi2_1}
    \chi^{2}(\phi_n,a_n) = \sum_{n,k} \frac{\lvert d_{nk} - a_{n}p_{nk}e^{-2\pi ik\phi_{n}}\rvert^{2}}{\sigma'^2_n}.
\end{equation}
It is useful at this point to make note of the fact that at a given frequency $\nu_n$, the above expression is equivalent to the FFTFIT prescription.  The fundamental difference in this approach, besides allowing for an arbitrary evolution of the pulse profile with frequency encoded in $p_{nk}$, is that we perform a global fit for both an achromatic phase $\phi^\circ_{ref}$ and a dispersion measure DM by implementing the constraint
\begin{equation}
\label{constraint}
    \phi_{n} = \phi^\circ_{ref} + \frac{K\times\textrm{DM}}{P_s}\Big(\nu_{n}^{-2}-\nu_{ref}^{-2}\Big),
\end{equation}
where $P_s$ is the pulsar's spin period and $\nu_{ref}$ is the dedispersion reference frequency.  This constraint reduces our minimization problem from having $2n_{chan}$ parameters, to $n_{chan} + 2$ (i.e. $\chi^{2}(\phi_n,a_n) \rightarrow \chi^{2}(\phi^\circ_{ref},\textrm{DM},a_n)$).  Ideally, we want to know what $\phi^\circ_{ref}$ is for $\nu_{ref} = \infty$.  However, we have chosen the above parameterization for the delays in each frequency channel, as opposed to the more specific infinite-frequency case of Equation~\ref{DMlaw}, because it allows us to find a reference frequency that gives zero covariance between the estimates of the phase and dispersion measure.  The form of the covariance between the estimates of $\phi^\circ_{ref}$ and $\textrm{DM}$ is given in the Appendix, which recommends that we choose $\nu_{ref}$ wisely (see also~\S\ref{mcs}).

By following a similar procedure as that written in~\citet{Demorest07}, and expanding and simplifying Equation~\ref{chi2_1} we obtain
\begin{equation}
\label{chi2_2}
    \chi^{2}(\phi^\circ_{ref},\textrm{DM},a_n) = S_d + \sum_{n}a_n^2S_{p,n} - 2\sum_{n}a_nC_{dp,n}
\end{equation}
where we have made use of the definitions
\begin{subequations}
\begin{equation}
\label{Sd}
    S_d \equiv \sum_{n,k} \frac{\lvert d_{nk}\rvert^{2}}{\sigma'^2_n},
\end{equation}
\begin{equation}
\label{Spn}
    S_{p,n} \equiv \sum_{k} \frac{\lvert p_{nk}\rvert^{2}}{\sigma'^2_n},
\end{equation}
and
\begin{equation}
\label{Cdpn}
    C_{dp,n}(\phi_n) \equiv \Re\Big\{\sum_{k} \frac{d_{nk}p_{nk}^*e^{2\pi ik\phi_{n}}}{\sigma'^2_n}\Big\}.
\end{equation}
\end{subequations}
The first two definitions are functions solely of the data and the model portraits.  If one considers discrete values of $\phi_n$ for a particular frequency channel $n$ ($\phi_{nj} = (j+0.5)/n_{bin}$), the third definition contains the inverse DFT of a multiplication of the data and the model, which is the same as the discrete cross-correlation of the time-domain quantities $D_{nj}$ and $P_{nj}$.  This definition highlights the fact that both FFTFIT and our extension of it across a discretized bandwidth can be thought of as cross-correlation techniques.

We can further simplify our minimization problem by recognizing that at the global minimum of the $\chi^2$ expression in Equation~\ref{chi2_2}, all of the first derivatives vanish.  Therefore, we only need to seek out a minimum of Equation~\ref{chi2_2} in the subspace where its partial derivatives with respect to all of the $a_n$ parameters are zero.  Solving for these maximum-likelihood $a_n$ as a function of the other parameters leads to the constraint
\begin{equation}
\label{an}
    a_{n} = \frac{C_{dp,n}}{S_{p,n}},
\end{equation}
which is inserted in Equation~\ref{chi2_2} to reduce our minimization problem to a two-parameter function,
\begin{equation}
\label{chi2}
    \chi^{2}(\phi^\circ_{ref},\textrm{DM}) = S_d - \sum_{n}\frac{C^2_{dp,n}}{S_{p,n}}.
\end{equation}
We retain the use of the label $\chi^2$ to emphasize that the above function is a subspace of Equation~\ref{chi2_2}, and shares the global minimum that we seek\footnote{The minimization of this quantity corresponds to maximizing the ``profile likelihood.''}.  In practice, one needs to maximize only the strictly positive second term in the above equation, since it contains all of the phase and dispersion information, and the first term is a constant function of the data.  It is easy to see that, for negligible profile evolution, if the dispersion measure is zero or, equivalently, the data have been correctly dedispersed for that observation's epoch, then this algorithm is akin to averaging TOAs obtained in the usual way using individually aligned templates.  However, if the pulsar's DM needs to be measured at every epoch, and profile evolution should be accounted for -- which are both likely true for most observations of MSPs with wideband receiver systems -- we claim that this is a natural extension to how TOAs are currently procured.  At the very least, this algorithm should perform no worse than traditional techniques.

We derive errors and covariances for the maximum-likelihood estimates of the parameters in the Appendix, but here we wish to underscore that it is possible to analytically determine a dedispersion reference frequency $\nu_{zero}$ that yields zero covariance between the estimates for $\phi^\circ_{ref}$ and DM, which is tested in~\S\ref{mcs}.  Lastly, we note that~\cite{Liu14} contemporaneously developed a very similar frequency-dependent TOA algorithm independent of our efforts, which may be employed as part of the European Pulsar Timing Array\footnote{\url{www.epta.eu.org/}} project~\citep{Kramer13}.

\subsection{Implementation}
\label{imp}

\subsubsection{Software}
\label{soft}

We have implemented our wideband timing algorithm in publicly-available \texttt{python} code\footnote{\url{www.github.com/pennucci/PulsePortraiture}}, which also includes a Gaussian-component-based portrait modeling routine, which is described below.  The code utilizes the \texttt{python} interface to the pulsar data analysis package \texttt{PSRCHIVE}\footnote{\url{www.psrchive.sourceforge.net/}}~\citep{Hotan04,vStrat12}, as well as recent versions of \texttt{numpy}\footnote{\url{www.numpy.org/}}, the optimization functions in \texttt{scipy}\footnote{\url{www.scipy.org/}}, and the non-linear least-squares minimization package \texttt{lmfit}\footnote{\url{www.newville.github.io/lmfit-py/}; \texttt{lmfit} is a Levenberg-Marquardt algorithm that we use for the modeling code.}.

The minimization of the function in Equation~\ref{chi2} is performed by a truncated Newton algorithm that comes packaged in \texttt{scipy}, and the off-pulse noise level is determined in PSRCHIVE by examining the variance in a window around a profile's minimum.  The initial phase parameter value is estimated by using a one-dimensional brute-force routine in \texttt{scipy}, which is performed on the frequency-averaged data and template.  In order to do this, the data are dedispersed with respect to an estimate for $\nu_{zero}$, which can only be determined after the minimum is found.  The nominal DM from the pulsar's ephemeris is used in this dedispersion and also as the initial DM parameter value in the global fit.  It is also possible to exclude fitting for a DM and only determine a phase.  The default behavior in the code transforms the best-fit phase estimate $\hat{\phi}^\circ_{ref}$ to reference $\nu_{zero}$, which gives the smallest, uncorrelated error for the TOA (see the Appendix).

It is important to use barycentric frequencies for $\nu_n$, otherwise the Earth's orbit induces an artificial yearly oscillation of the DM from the Doppler-shifted frequencies.  Alternatively, one can simply propagate the Doppler factor $\Gamma$ through the frequency and temporal terms of Equation~\ref{DMlaw} to correct the observed ``topocentric'' dispersion measure DM$_{topo}$,
\begin{equation}
\label{doppler}
    %\textrm{DM} = \frac{\textrm{DM}_{topo}}{\Gamma},  # took years to find and flip this...
    \textrm{DM} = \Gamma \times \textrm{DM}_{topo}\footnote{NB: this equation was erroneously inverted in earlier versions of this paper, as well as the published ApJ version, but was always implemented correctly in the code.},
\end{equation}
where
\begin{equation}
\label{Gamma}
    \Gamma \equiv \sqrt{\frac{1+\beta}{1-\beta}},
\end{equation}
\begin{equation}
\label{beta}
    \beta \equiv \frac{v}{c},
\end{equation}
and $v$, the projected velocity of the observatory onto the line-of-sight, is positive for growing separation.  With respect to the demonstration in the next section, our source is close to the ecliptic plane and has a large DM, so this correction was essential.

\subsubsection{Portrait Modeling}
\label{modeling}

It is obvious that there is freedom in the choice of model portrait to use and we stress that any arbitrary model can be used in the above algorithm for phase and DM measurements.  We have experimented almost exclusively with analytic Gaussian-component models, but it is also feasible to find an interpolation scheme based on an average of all the data portraits (or, for example, a principal component analysis approach).  However, Gaussian-component modeling has been used extensively in the literature (for instance, see~\citet{Foster91, Kramer98, Lommen01, Ahuja07, Hassall12}) and is a simple way to generate analytic noise-free templates.  We model pulse profile evolution with independently changing Gaussian components $g_i$ of the form
\begin{equation}
\label{Pnu}
    P(\nu, \varphi) = \sum_i g_i(\nu, \varphi),
\end{equation}
where
\begin{equation}
\label{gi}
    g_i(\nu, \varphi|A_i, \varphi_i, \sigma_i) = A_i(\nu)\textrm{exp}\big(-4\textrm{ln}(2)\frac{(\varphi-\varphi_i({\nu}))^2}{\sigma_i(\nu)^2}\big).
\end{equation}
We choose to model the positions $\varphi_i$, widths (FWHM) $\sigma_i$, and amplitudes $A_i$ as power-law functions of frequency,
\begin{equation}
\label{PL_comps}
    X_i(\nu | X_{\circ,i}, \alpha_{X,i}, \nu_{\circ}) = X_{\circ,i}\Big(\frac{\nu}{\nu_{\circ}}\Big)^{\alpha_{X,i}},
\end{equation}
for Gaussian parameter $X$ and model reference frequency $\nu_{\circ}$.  We also include linear functions for $\varphi_i$ and $\sigma_i$ in the code.  The modeling code allows flexibility for any of the parameters to be fixed; for example, a ``fiducial component'' with no positional change as a function of frequency can be selected.  Most MSP portraits we have experimented on seem to be sufficiently characterized by a few to roughly a dozen or so Gaussian components.  We also include an option to include scattering in the fit for the model via a convolution with a one-sided exponential,
\begin{equation}
\label{Pscat}
    P(\nu, \varphi) = P_{unscattered}(\nu, \varphi) \ast e^{-\frac{\varphi P_s}{\tau(\nu)}}H(\varphi),
\end{equation}
where
\begin{equation}
\label{taunu}
    \tau(\nu) = \tau_{\circ}\Big(\frac{\nu}{\nu_{\circ}}\Big)^{\alpha_{scat}}
\end{equation}
is the scattering timescale, $H$ is the Heaviside step function, and we have assumed $\alpha_{scat} = -4.0$~\citep{Bhat04}.  One could imagine extending our algorithm to include a variable scattering parameter in the fit to the data, instead of fixing it in the model.  The benefits, applicability, and practical limitations of doing this are currently being investigated by the authors.  The details of pulse portrait modeling and its physical interpretations are beyond the scope of this paper, but we demonstrate one application of Gaussian modeling in the next section.

Finally, one subtlety that we did not address is the averaging of the model within each frequency channel to match the channel bandwidth of the data.  Insofar that the aim is to have a greater number of channels, this should have a negligible effect.  Presumably, each channel's profile evolution is minute and, as given in Equation~\ref{DMsmear}, the channel smearing from an inaccurate DM is also small.  Similarly, we assumed that a sufficient number of phase bins are used so that all of the harmonic content of the profile is retained and an averaging of the model into phase bins is well-approximated by the Gaussian model evaluated at the phase bin centers.  However, a more rigorous representation for $P_{nj}$ would be one that multiplies the model by a sampling function that has been convolved with both the channel-width and bin-size of the data.

\section{Demonstration with MSP J1824-2452A}
\label{m28a}

\subsection{M28A Dataset and Model Portrait}
\label{m28a_data}

Pulsar J1824-2452A (M28A, hereafter) is a highly energetic, bright, isolated 3.05~ms pulsar in the globular cluster M28~\citep{Lyne87, Johnson13}.  We chose this MSP as a demonstrative case-study because it has a large dispersion measure ($\approx$120~cm$^{-3}$~pc), a large DM gradient (several $\times10^{-3}$~cm$^{-3}$~pc~yr$^{-1}$)~\citep{Backer93, Cognard97, Keith13}, a complex profile with broad and narrow features, and because it shows component evolution across the frequency range 720 -- 2400~MHz~\citep{Foster91}.

\begin{deluxetable}{lccrc}
\centering
\tablewidth{0pt}
\tablecaption{Observed Epochs and Measured DMs of M28A with GUPPI\label{M28AT}}
\tablehead{\colhead{Epoch}   &  \colhead{MJD}   &   \colhead{$\nu_c$}   &   \colhead{Length}    &   \colhead{$\Delta$DM} \\
\colhead{[UTC]}  &  \colhead{[day]} &   \colhead{[MHz]} &   \colhead{[min]}  &   \colhead{[$\times10^{-3}$~cm$^{-3}$~pc]}}
\startdata
2010-02-11	&  55238.72 	&	1500	&	43.3$\ \ $  &	$-2.4  \pm 0.2$    \\ 
2010-05-20*	&  55336.35 	&	2000	&	129.0$\ \ $ &	$-2.2  \pm 0.4$    \\ 
2010-08-11	&  55419.15 	&	2000	&	166.3$\ \ $ &	$-0.6  \pm 0.4$    \\ 
2010-10-05*$\dagger$	&  55474.00 	&	820	    &	159.4$\ \ $ &	$ \ \ 0.13  \pm 0.07$   \\ 
2010-10-20*	&  55489.93 	&	1500	&	149.3$\ \ $ &	$ \ \ 0.18  \pm 0.06$   \\ 
2011-03-05	&  55625.58 	&	1500	&	157.2$\ \ $ &	$ \ \ 2.58  \pm 0.08$   \\ 
2011-04-04	&  55655.55 	&	1500	&	98.8$\ \ $  &	$ \ \ 1.56  \pm 0.09$   \\
2011-04-13*	&  55664.42 	&	1500	&	154.2$\ \ $ &	$ \ \ 1.42  \pm 0.06$   \\ 
2011-07-02*	&  55744.25 	&	1500	&	149.3$\ \ $ &	$-0.12 \pm 0.06$   \\ 
2011-09-29	&  55833.98 	&	1500	&	154.4$\ \ $ &	$ \ \ 0.28  \pm 0.06$   \\ 
2012-01-06	&  55932.72 	&	1500	&	145.3$\ \ $ &	$-1.91 \pm 0.06$   \\ 
2012-04-09*	&  56026.45 	&	1500	&	149.3$\ \ $ &	$-2.81 \pm 0.04$   \\ 
2012-04-15*	&  56032.42 	&	2000	&	161.4$\ \ $ &	$-2.8  \pm 0.2$    \\ 
2012-07-03	&  56111.25 	&	1500	&	138.3$\ \ $ &	$-2.23 \pm 0.06$   \\ 
2012-10-07$\dagger$	&  56207.96 	&	1500	&	152.3$\ \ $ &	$-0.55  \pm 0.06$   \\ 
2013-01-06	&  56298.70 	&	1500	&	149.3$\ \ $ &	$ \ \ 1.54  \pm 0.06$   \\ 
2013-04-08*	&  56390.48 	&	1500	&	176.4$\ \ $ &	$ \ \ 0.02  \pm 0.05$   \\ 
2013-04-15*$\dagger$	&  56397.46 	&	2000	&	164.4$\ \ $ &	$ \ \ 0.1    \pm 0.3$    \\ 
2013-05-06	&  56418.29 	&	2000	&	82.2$\ \ $  &	$ \ \ 1.0   \pm 0.5$    \\
2013-05-09	&  56421.44 	&	2000	&	83.2$\ \ $  &	$ \ \ 1.5   \pm 0.7$    \\
2013-05-11*	&  56423.40 	&	2000	&	77.2$\ \ $  &	$ \ \ \ 0.8   \pm 0.5$    \\
2013-05-13*	&  56425.43 	&	2000	&	82.2$\ \ $  &	$ \ \ \ 1.4   \pm 0.5$    \\
2013-05-18	&  56430.27 	&	2000	&	83.2$\ \ $  &	$ \ \ \ 0.7   \pm 0.5$    \\
2013-05-24	&  56436.43 	&	2000	&	59.1$\ \ $  &	$ \ \ \ 1.6   \pm 0.7$    \\
2013-05-31	&  56443.18 	&	2000	&	80.2$\ \ $  &	$ \ \ \ 0.6   \pm 0.4$    \\
\enddata
\tablecomments{The columns are the UTC YYYY-MM-DD observation date, the Modified Julian Date, the center frequency, the total integration time, and the measured DM with 1$\sigma$~uncertainties.  The DMs had the nominal (unweighted) average value of 119.88818~cm$^{-3}$~pc subtracted.  There are 11~epochs observed at 2000~MHz (800~MHz~BW), 13~at 1500~MHz (800~MHz~BW), and 1~at 820~MHz (200~MHz~BW).  The fractional bandwidths are approximately 0.25, 0.53, and 0.40 for the 820, 1500, and 2000~MHz data, respectively.  The starred epochs were used in the fit for the Gaussian model, and epochs with a dagger are shown as part of Figure~\ref{M28A_freqevol_compare}.}
\label{epochs}
\end{deluxetable}

The M28A dataset presented here consists of 25 epochs of multi-frequency observations spanning more than three years from the Green Bank Telescope.  The data were obtained with GUPPI beginning in February 2010 soon after the implementation of its real-time coherent dedispersion capability, which was utilized for the taking of these observations in search-mode (i.e as unfolded time-series).  Each of the time-series for the 512 channels across each frequency band were dedispersed at the nominal average DM for the globular cluster, 120~cm$^{-3}$~pc, and then folded using a predetermined ephemeris for M28A.  The native resolution of the data is $10.24~\mu$s, which is sufficient to resolve the profile, although we folded the data at nearly twice this resolution, resulting in 512 phase bins.  A more technical description of these data and their calibration will be provided in a forthcoming paper by Bilous et al.  Table~\ref{epochs} summarizes the epochs of the observations presented here.

\begin{figure*}
\epsscale{1.0}
\plotone{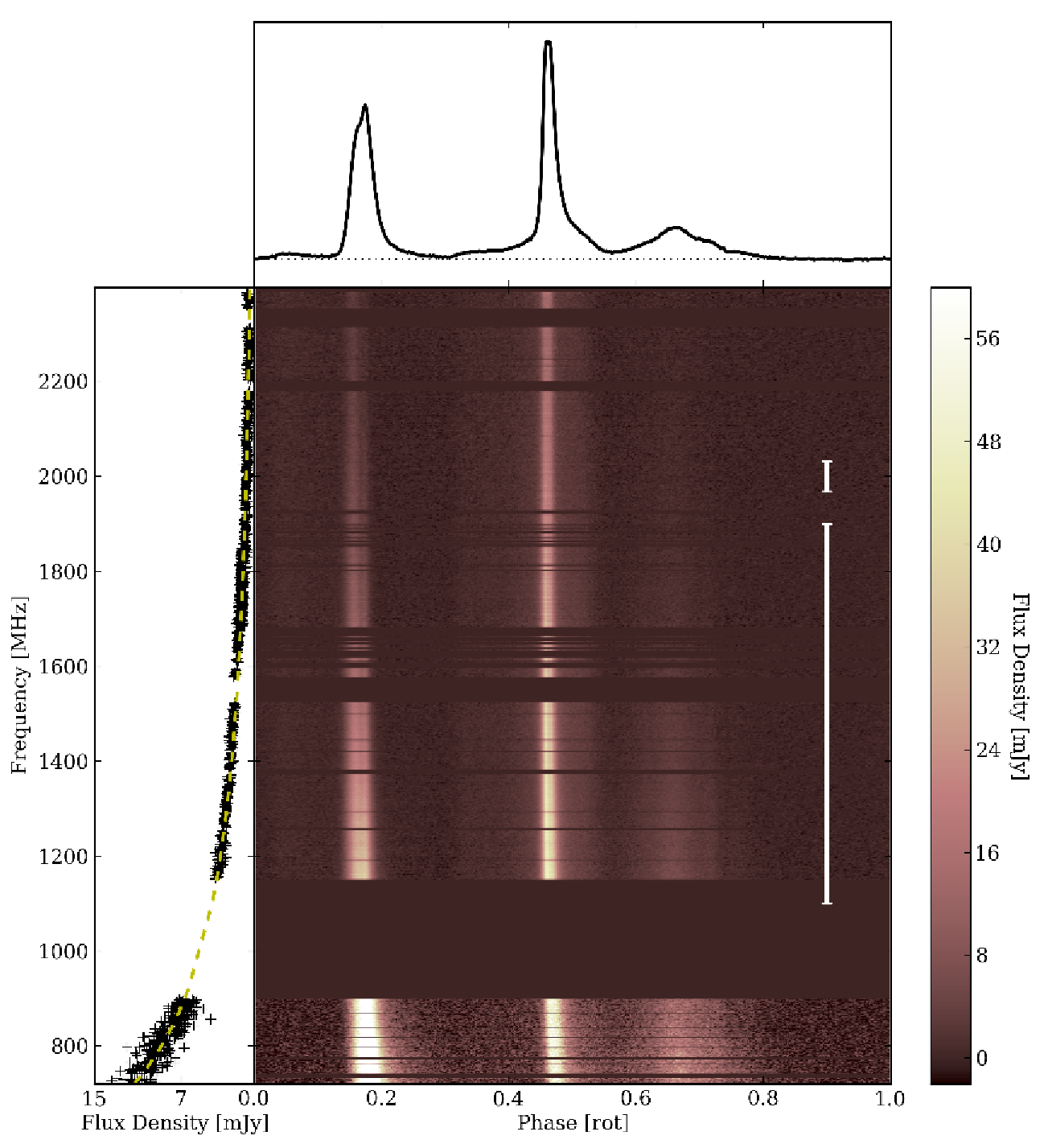}
\caption{{A portrait of concatenated M28A data from time-averaged observations taken at 820~MHz, 1500~MHz, and 2000~MHz.  See Table~\ref{epochs} for the epochs used in the averages and the text for a complete description.  The horizontal gaps are where radio-frequency interference was excised, but the widest one is a coverage gap between receivers.  The two white vertical bars show the bandwidth coverage offered by previous coherent dedispersion backends (64~MHz) and current ones like GUPPI (800~MHz).  Note that the UBB has an instantaneous bandwidth of almost 1.5 times that shown above.}}
\label{M28A_port}
\end{figure*}

Figure~\ref{M28A_port} shows a concatenated portrait of several epochs of the M28A data, displaying an effective bandwidth of $\sim$1.5~GHz.  The complexity of the portrait is evinced by its asymmetries, its non-Gaussian features, the exchanging dominance of components from differing spectral indices, and the presence of an obvious scattering tail at the lower frequencies.  To make this portrait, five high signal-to-noise ratio (SNR) epochs were selected from each set of 1500~MHz and 2000~MHz observations, they were each averaged together based on the ephemeris, and then joined in tandem along with the 820~MHz observation in a fit for the two-dimensional Gaussian model, as described in~\S\ref{modeling}.  The fit included nuisance phase and DM parameters for each band, as well as a scattering timescale $\tau_{\circ}$.  In effect, the nuisance parameters attempted to ``align'' the data so that the Gaussian parameters can be optimized.

\begin{deluxetable}{cc@{ }rc@{   }rcc}
\centering
\tabletypesize{}
\tablewidth{0pt}
\tablecaption{Analytic Gaussian Model Parameters for M28A}
\tablehead{\colhead{$i$}   &   \colhead{$\varphi_{\circ,i}$}   &  \colhead{$\alpha_{\varphi,i}$}   &  \colhead{$\sigma_{\circ,i}$}    &   \colhead{$\alpha_{\sigma,i}$}  &   \colhead{$A_{\circ,i}$} &   \colhead{$\alpha_{A,i}$} \\
\colhead{}  &   \colhead{{\small[rot]}}  &  \colhead{}  & \colhead{{\small[\% rot]}}  &    \colhead{}  &   \colhead{}  &  \colhead{}}
\startdata
1  & -0.00180$ \ $  &  -0.693  &  10.00  &  0.3  &  0.09  &  -1.2 \\
2  &  0.00000  &   0.000  &   0.88  &  0.2  &  1.11  &  -1.9 \\
3  &  0.00410  &  -0.021  &   2.24  &  0.3  &  0.63  &  -2.1 \\
4  &  0.00932  &  -0.017  &   0.69  & -0.1  &  0.58  &  -1.4 \\
5  &  0.02078  &   0.280  &   9.96  & -2.0  &  0.13  &  -0.1 \\
6  &  0.18894  &  -0.006  &   7.93  &  0.4  &  0.08  &  -3.6 \\
7  &  0.21877  &  -0.124  &  10.00  &  0.0  &  0.05  &  -3.3 \\
8  &  0.70012  &  -0.007  &   2.24  & -0.1  &  0.75  &  -3.0 \\
9  &  0.71061  &  -0.025  &   9.98  &  5.3  &  0.05  &  -6.1 \\
10 &  0.71651  &  -0.001  &   1.09  &  0.2  &  0.42  &  -3.5 \\
\hline
$\tau_{\circ}$ &   4.57~$\mu$s$\ $   &   &   &   &   &   \\
$\nu_{\circ}$  &   1500.00~MHz$\ \ $   &   &   &   &   &   \\
\enddata
\tablecomments{The column headers are defined in Equation~\ref{PL_comps}.  The components are ordered by phase; Figures~\ref{M28A_port}, \ref{M28A_resids}, and~\ref{M28A_scint_demo} have been rotated for clarity.  The second component listed is the ``fiducial component''.  A limit of 0.1 rotations was placed on the FWHM width of the components to prevent runaway for small-amplitude components.  The precision of all the parameters is arbitrary, since we offer no interpretation of the model in this paper.  The reference frequency for the toy model is 1500~MHz and a scattering kernel corresponding to a fitted scattering timescale of $\tau_{\circ} \approx 5\ \mu$s at 1500~MHz was applied to the model (cf. Equations \ref{Pscat} and \ref{taunu}).  The point estimate of the scattering timescale is marginally consistent with that found independently from a separate analysis of giant pulses in this M28A data (Bilous, private communication).}
\label{model}
\end{deluxetable}

\begin{figure*}
\epsscale{1.0}
\plotone{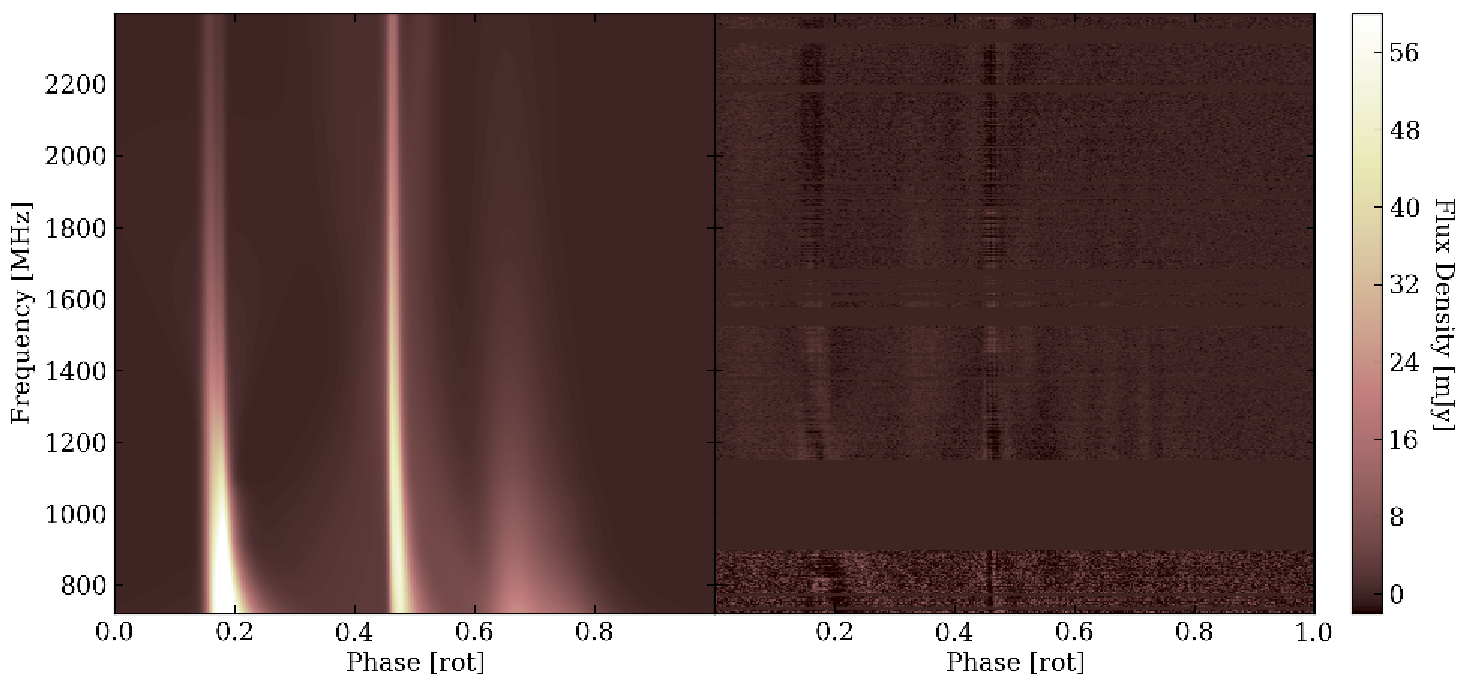}
\caption{The constructed Gaussian model and residuals after subtracting the data in Figure~\ref{M28A_port}.  We used a ten-component model that captures both the finer structure seen at the higher frequencies and the scattering at the lower frequencies.  Although the model and residuals show that the Gaussian modeling is not perfect, the model still proved sensible for timing and DM measurements.  The phase-averaged spectral index of the model is consistent with that of Figure~\ref{M28A_port}.}
\label{M28A_resids}
\end{figure*}

\begin{figure}
\epsscale{1.0}
\plotone{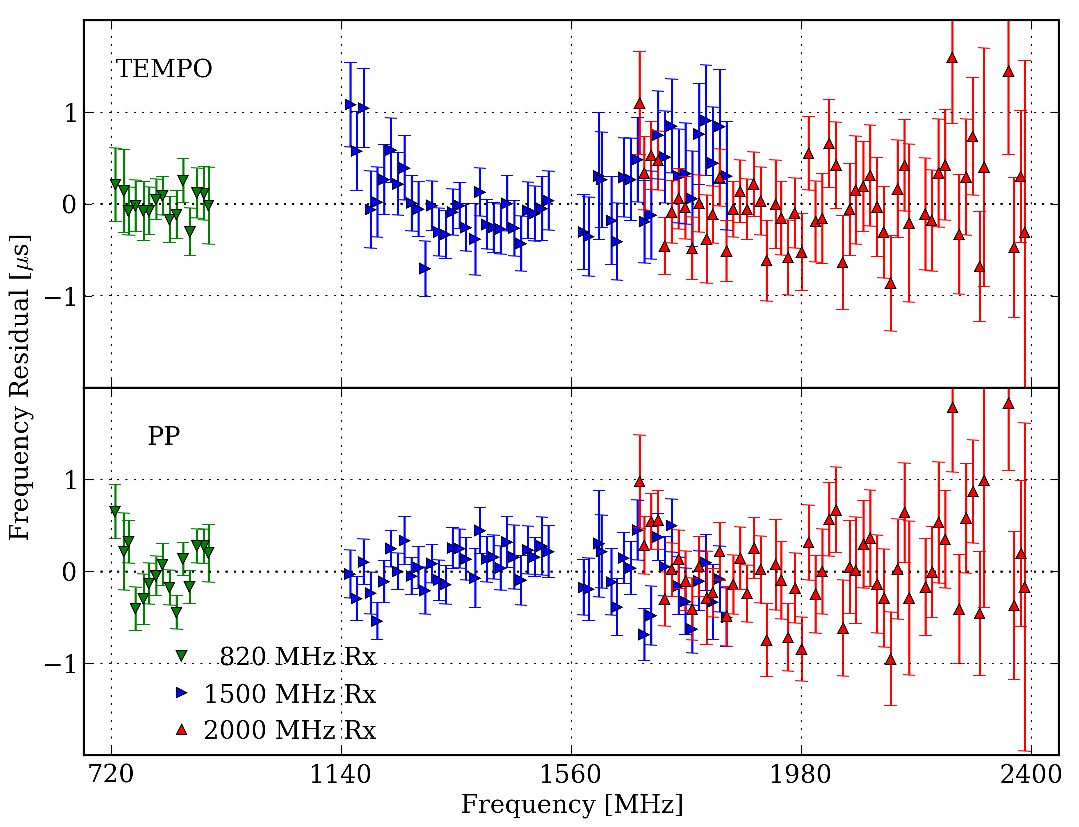}
\caption{Timing residual structure demonstrating profile evolution as a function of frequency for some M28A data after the DMs have been measured (see~\S\ref{compare} for discussion).  Table~\ref{epochs} denotes which epochs are shown.  Each point represents 12.5~MHz of bandwidth averaged.  The top panel (``TEMPO'') does not account for any profile evolution, but only uses a single template profile per receiver band, producing the observed scatter.  The reduced $\chi^2$ for the single 820 MHz observation shown was anomalously low at $\sim0.3$.  The opposing trends in the overlapping region between the 1500 and 2000~MHz residuals signify that there is no continuous frequency-dependent model.  A comprehensive, global model for the evolution would ideally show flat residuals.  The ``residuals'' from applying our algorithm with the Gaussian toy-model are shown in the lower panel (``PP''), showing the best results in the 1500~MHz data.  Next generation wideband receivers will simultaneously cover more than this entire spectrum at once.}
\label{M28A_freqevol_compare}
\end{figure}

Following the suggestion in~\citet{Foster91}, we modeled the widths of M28A's components with power-law functions, and had less success when trying linear models.  To obtain initial parameters for the two-dimensional model, we fit ten Gaussian components to a profile referenced at 1500~MHz, representing 200~MHz of bandwidth averaged.  We chose the dominant component at 1500~MHz to be a fixed ``fiducial component''.  The parameters of the fitted model are given in Table~\ref{model}.  MSPs like M28A exemplify how the choice of a ``fiducial point'' is not simple (eg. see~\citet{Craft70}) because the profile has no obvious symmetries, and the dominant component changes as a function of frequency.  The model and the residuals are shown in Figure~\ref{M28A_resids}.

The thick solid black line in the top panel of Figure~\ref{M28A_port} represents the frequency-averaged light curve of the aligned data.  This profile marginalizes over all of the frequency structure and scattering tails, and so it would be imprudent to use such a profile as a template for obtaining phase measurements.  The phase-averaged spectral flux density profile in the left panel was fit with a power-law (yellow dashed line).  We obtained a spectral index of $-2.36 \pm 0.02$, although it appears as though the flux is not perfectly modeled by a single power-law.  Details of M28A's spectra from this dataset will also be presented in a forthcoming paper (Bilous et al., in preparation).  

Some of the subtle profile evolution for this pulsar can be seen in the top panel of Figure~\ref{M28A_freqevol_compare}, which consists of timing residuals as a function of frequency (see~\S\ref{compare}).  As is evident from the top panel, and particularly in the 1500~MHz data, the use of an average template profile to measure TOAs for a band, or portion thereof, will produce a different residual as a function of frequency based on the profile's departure from the frequency-averaged template.  If this frequency-dependent bias were constant, in would be absorbed into the timing model, but varying scintillation patterns can change which segments of the bias are weighted more significantly (or, similarly, what the frequency-averaged profile looks like), thereby introducing random systematic noise into the timing residuals.

\begin{figure*}
\epsscale{1.0}
\plotone{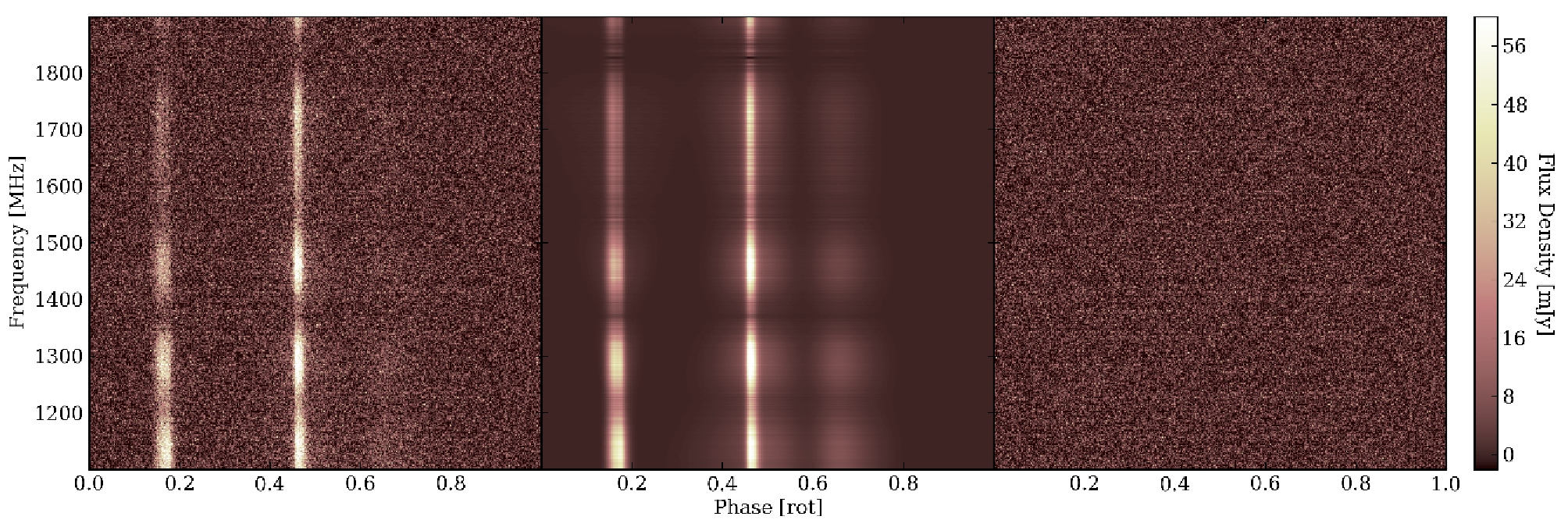}
\caption{A simple demonstration of how the algorithm automatically manages diffractive scintillation.  A random phase, DM, and ``scintillation pattern'' was added to the 1500~MHz portion of the model in Figure~\ref{M28A_resids}, along with frequency-independent noise (left panel), and then a fit was performed.  The fitted model is shown in the middle panel, which encodes the scintillation in the $a_n$ scale parameters, and the fit residuals are to the right.  The residual statistics match the off-pulse noise and mean from the input data.  A fake-data sample of Medium SNR from~\S\ref{mcs} has about the same SNR as these data.}
\label{M28A_scint_demo}
\end{figure*}

The scintillation bandwidth for M28A ($\sim$0.016~MHz at 1~GHz~\citep{Foster91}) is much smaller than any of the observed channel bandwidths, so the data do not show obvious scintles in the folded profiles.  However, to demonstrate the utility of fitting for the $a_n$ parameters ``for~free'', Figure~\ref{M28A_scint_demo} shows an example fit to fake data of moderate SNR generated by adding a fake ``scintillation pattern'' and frequency-independent noise to the M28A model in Table~\ref{model}.  A random phase and $\Delta$DM was added to the data, and then it was run through our code, producing the fitted model and residuals shown in the figure.  The fitted $a_n$ values provide information about diffractive scintillation from the ISM, and they effectively act as weights for individual multi-channel TOAs that have been fitted for a DM and averaged together to obtain $\phi^{\circ}_{ref}$.  Ideally, this advantage obviates the need to cull very low SNR TOAs of individual frequency channels.  In principle, the $a_n$ values could also be used to determine if there is residual RFI in the data, although we have not yet investigated how the presence of RFI will affect the fitting.

\begin{figure}
\epsscale{1.0}
\plotone{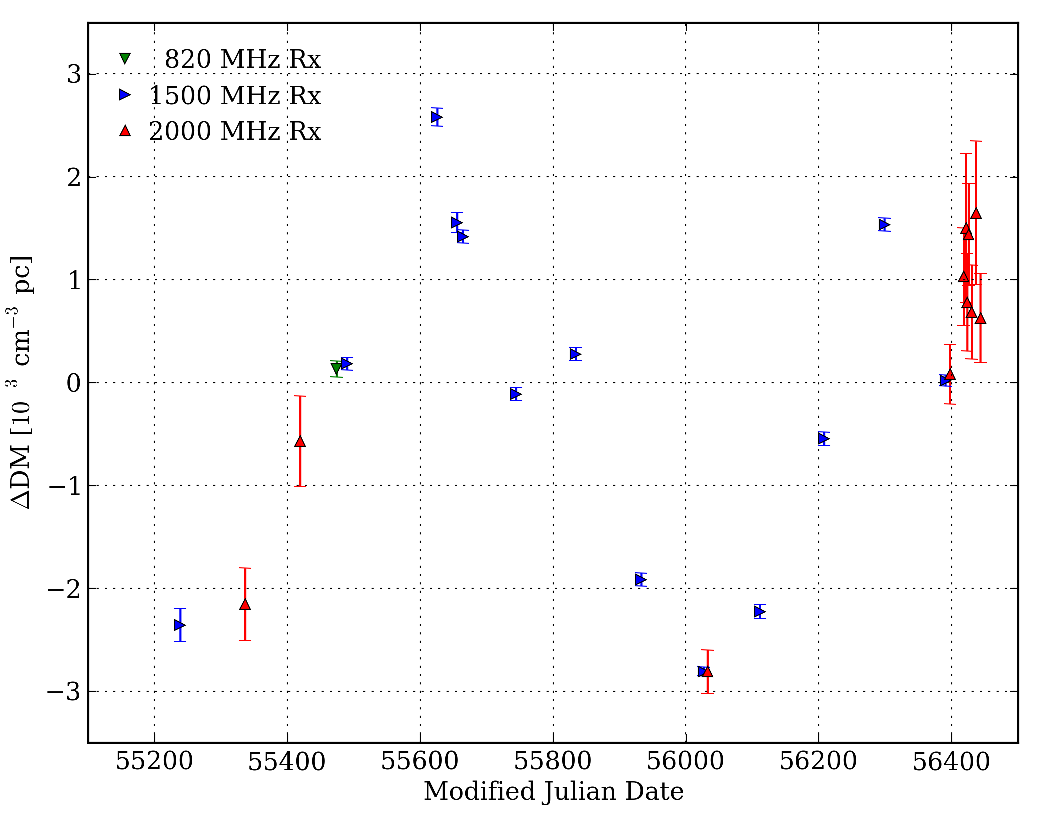}
\caption{The twenty-five DM measurements from Table~\ref{epochs}.  The calendar range of the data spans Feb 11, 2010 to May 31, 2013.  \citet{Keith13} reports a similar $\sim$5 $\times 10^{-3}$~cm$^{-3}$~pc increase in the first third of our data.}
\label{M28A_DMt}
\end{figure}

Using the algorithm described in~\S\ref{desc}, the Gaussian model was used as~$P_{nk}$ to fit for TOAs and DMs in the twenty-five observed epochs~\{$D_{nk}$\}.  The average per-epoch TOA uncertainty is $\sim$40~ns in the 1500~MHz data, and $\sim$90~ns in the 2000~MHz data.  Figure~\ref{M28A_DMt} shows the measured DM variations for the M28A dataset, where an average DM of 119.88818~cm$^{-3}$~pc was subtracted.  We obtained DM precisions between several~$\times 10^{-5}$ and several~$\times 10^{-4}$~cm$^{-3}$~pc.  For the 1500~MHz data, the average DM precision of $\sim$7~$\times 10^{-5}$~cm$^{-3}$~pc corresponds to about 160~ns~$\approx 5 \times 10^{-5}$~rotations~$\approx$~0.03~bin of drift across the band, for 512 phase bins.  It is interesting to compare this number to the amount of dispersive smearing in each channel from coherently dedispersing these data with the incorrect DM of 120~cm$^{-3}$~pc; at 1500~MHz, $\delta t_{\textrm{DM}} \approx 430$~ns.  The first third of our measurements overlap with observations of M28A presented in~\citet{Keith13}; the overall trend in our DM measurements in these epochs is consistent with what is seen in their data.

\subsection{Comparison of Methods}
\label{compare}

In what follows, we have compared our measurements (labeled ``PP'') with those obtained from the same data using a more traditional procedure (which we collectively label ``TEMPO'').  For the latter, multi-channel TOAs were obtained via standard techniques: each time-averaged epoch's band was divided into 64 channels, and each channel's profile was cross-correlated with a smoothed template profile that was obtained by averaging all the data from a given receiver.  An FFTFIT-based algorithm was used for the pulse phase fitting (the cross-correlation)\footnote{Specifically, we used the Fourier phase gradient (PGS) algorithm in the \texttt{PSRCHIVE} program \texttt{pat}.}.  Each epoch's DM was then determined by individually fitting a {\it fixed} timing model to the epoch's TOAs using the popular pulsar timing software \texttt{tempo}\footnote{\url{www.sourceforge.net/projects/tempo/}}, {\it allowing only the dispersion measure to vary}\footnote{For clarity, at no time did we do a multi-frequency or multi-epoch fit for DM, although this is one area of current research.}.  In effect, this process removes a quadratic delay across the multi-channel TOAs.  No consideration of profile evolution is taken into account besides the usage of three separate template profiles for the three bands.  Therefore, if all of the TOAs were used in a timing model fit, the use of arbitrary phase-offsets (JUMPs) between the three sets of TOAs would be needed to align the template profiles.

\subsubsection{Mitigation of Profile Evolution}
\label{mitigation}

The top panel of Figure~\ref{M28A_freqevol_compare} shows typical \texttt{tempo} multi-channel frequency residuals from not modeling the profile evolution.  Note that the 820~MHz data is shown here to have the same channel bandwidth as the higher frequencies.  Introducing phase-offsets to align small portions of the band (or from using numerous templates) is one approach to remove the frequency-dependent structure, but it adds a large number of otherwise meaningless parameters into the timing model \citep{Demorest13}.

A somewhat less arbitrary approach is to characterize the trend with a simple function that can be included in the timing model.  This latter multi-channel TOA strategy, combined with a \texttt{tempo} fit for the profile evolution and variable DM, is akin to the current timing methodology employed by the PTA collaboration called NANOGrav\footnote{The North American Nanohertz Observatory for Gravitational Waves: \url{www.nanograv.org}}$^{\textrm{,}}$\footnote{Here, we are referring to the use of a \texttt{tempo} functionality called ``DMX'', plus a polynomial function of log-frequency to account for profile evolution.  With DMX, the discrete DMs are measured in situ with other timing model parameters while using the overall WRMS residual as the discriminating quantity, which means that the DMs and profile evolution parameter can absorb unmodeled, non-ISM effects like timing noise, or a gravitational wave signal.}~\citep{McLaughlin13}.  Ignoring profile evolution altogether and using frequency-averaged profiles may still be a sufficient practice for particular pulsars.  However, this strategy will become untenable with the next generation of receivers.  It seems more appropriate and simple to model directly the profile evolution based on the folded profiles, as we have done, and then simultaneously measure a TOA and a DM.

For the sake of comparison, we made multi-channel residuals {\it after} applying our algorithm to each of the same epochs in Figure~\ref{M28A_freqevol_compare}, and have plotted them in the lower panel.  These ``residuals'' were calculated by {\it independently} cross-correlating each channel in the {\it fitted} two-dimensional model with the corresponding frequency channel in the data portrait using our own FFTFIT routine.  The greatest improvement in modeling the profile evolution is seen in the 1500~MHz data, and we will show several consequences of this in the following sections.  This improvement is sensible because the 1500~MHz data is our best ``wideband'' data in that it has the largest SNR, the largest fractional bandwidth, and hence the most profile evolution to be characterized.  There is also continuity in the residuals with the 2000~MHz data (from a separate epoch).  The 2000~MHz data remains qualitatively the same because of its lower SNR and smaller fractional bandwidth (i.e. less observable profile evolution).  The scatter of both sets of points is about the same as the corresponding average residual uncertainty.

On the other hand, the slight arch that remains in the 1500-MHz residuals and the added scatter into the 820-MHz residuals highlight the insufficiencies of a simple Gaussian modeling scheme for such a complex profile.   The scatter in the 820~MHz points may be explained by the difficulty of characterizing its simpler profile with too many evolving Gaussian components, including scattering, although the fact that we did not apply any averaging to the model within each 12.5~MHz-wide channel may also play a role.  Alternatively, our attempt to concatenate disparate bands from epochs that may have significantly different scattering timescales may have negatively affected the lowest frequencies.  More simply, an evolving Gaussian-component model does not describe the data well across all of the observed frequencies, but having more data in the 180~MHz-wide gap between the 820~MHz and 1500~MHz bands could help us find a better model.  Note that these residuals are specific examples from the whole dataset, and that the goodness of the model's fit will vary from observation to observation.

\subsubsection{Comparison of Dispersion Measures}
\label{DMs}

\begin{figure}
\epsscale{1.0}
\plotone{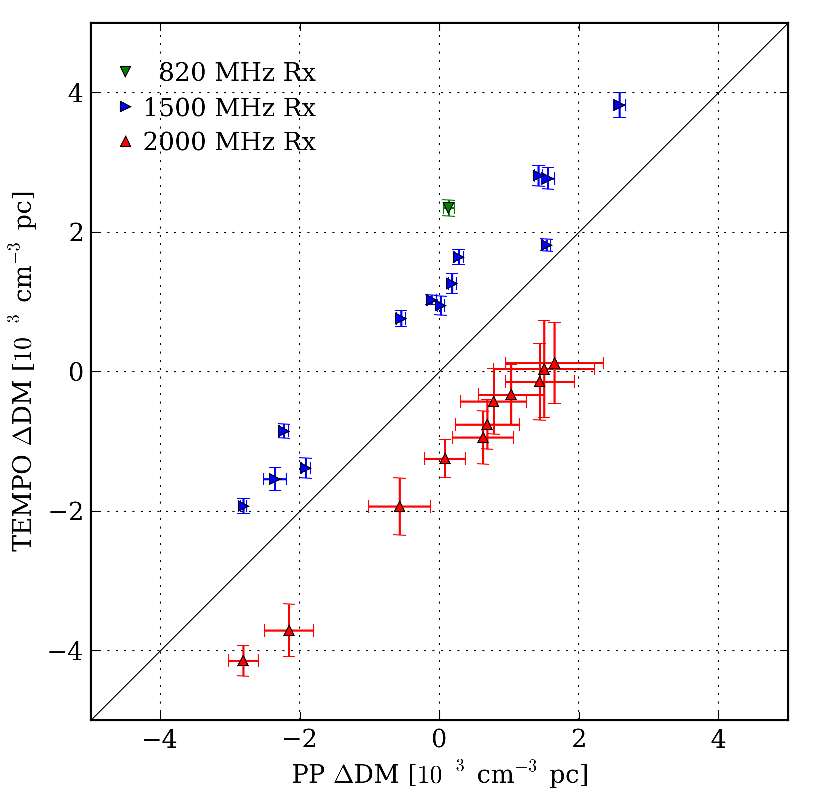}
\caption{A comparison of mean-subtracted DM trends as measured in the M28A data by our technique and a more usual approach.  ``PP'' represents our measurements.  The solid line traces equality.  The error bars for the ``TEMPO'' DMs are given by the least-squares \texttt{tempo} fit and have been multiplied by the reduced-$\chi^2$ value, which only significantly affected the 1500~MHz points at the level of tens of percent.  These are the same uncertainties as those shown in Figure~\ref{M28A_error_compare}.  The calculation of the error on our DM measurements is provided in the Appendix.  The $\sim$2 $\times 10^{-3}$~cm$^{-3}$~pc offset arises from the difference in how profile evolution is modeled.  See text for further discussion.}
\label{M28A_DM_compare}
\end{figure}

The absolute DM is not a useful measure for comparison because its values depend on how the profile and its evolution are modeled, and the DM can even vary based on its inclusion in a timing model fit (not applicable here).  Consequently, the average values of differently measured DMs will differ by a constant.  Figure~\ref{M28A_DM_compare} shows that our mean-subtracted DMs are in agreement with those obtained from the above described methods.  That is, the DMs measured in the 1500~MHz and 2000~MHz epochs are parallel to the solid line that represents equality, and so the two sets of measurements track roughly the same changes in DM as a function of time.

The strong agreement in the DMs from the 2000~MHz data corroborates with our statement above about the similarity of the multi-channel residuals; these $\Delta$DMs agree within their errors and have a scatter of $\lesssim2 \times 10^{-4}$~cm$^{-3}$~pc~$\approx 180$~ns of drift across the band.  In a similar vein, the observed scatter in the 1500~MHz data implies that our mitigation of the profile evolution mentioned above has significantly altered the measured DM trend.  Here, the $\Delta$DMs are scattered by $\lesssim4 \times 10^{-4}$~cm$^{-3}$~pc~$\approx 900$~ns of drift (with the largest deviation at three times that level).  These significant differences may be relevant when interpolating the behavior of the ISM between observations.  We address the measurement uncertainties in the next section.

The offsets seen in the figure between the DMs measured in each receiver band comes from the different modeling of profile evolution in each band.  For example, the ``TEMPO'' DMs are measured with three different templates that are assumed to be constant as a function of frequency in their respective band.  If this is a better assumption at higher frequencies, then the apparent average DM will be a function of frequency.  Indeed, the pairs of observations that were separated by only $\sim$1~week and taken at different frequencies show an offset of $\sim2 \times 10^{-3}$~cm$^{-3}$~pc, which is much larger than any of the differences between DMs measured in the same band, on the same time scale.

Similarly, having tried a vast number of fitted Gaussian models for M28A, we found that switching between different families of models produced the same large offsets (up to a few $\times 10^{-3}$~cm$^{-3}$~pc) between the DMs measured in different bands.  Given the reasonable assumptions we made about our model, we believe that the large frequency-dependent offsets seen in the DMs measured by using other fitted models (and the ``TEMPO'' templates) is explained by a misrepresentation of M28A's profile evolution and not, for instance, a frequency-dependent DM.  In fact, we used the assumption that no such offset exists in temporally proximate data from different frequencies as a qualitative model-selection criterion.

Ultimately, dispersion measures will be a function of frequency since the multi-path propagation of different frequencies will sample slighty different total free-electron column densities.  However, there is ambiguity between a frequency-dependent DM and profile evolution; as noted by~\citet{Ahuja05, Ahuja07},~\citet{Hassall12}, and others, an apparent frequency-dependent DM can be explained by unmodeled profile evolution.  For example, again consider Figure~\ref{M28A_freqevol_compare}; the DM measured using different sets of frequencies would vary because the phase-offset between arbitrarily chosen pairs of frequencies is not constant.

Potentially, in a bright, highly-scattered, high-DM pulsar like M28A, a frequency-dependent DM could be detected.  A rough estimate of the level of $\delta$DM($\nu$) that could be expected in the data can be estimated from M28A's scattering measure and distance as reported in~\citet{Foster91} in combination with the prediction for the form of a frequency-dependent DM in~\S4.4 of~\citet{CordShan10}.  However, the prediction is nearly proportional to the unknown distance to the scattering material.  Furthermore, a constant offset between DMs determined in different frequency bands is highly covariant with profile shape evolution, as described above.

It may become feasible to disentangle a frequency-dependent DM from profile evolution when truly broadband (eg. fractional bandwidth $\gtrsim$1), long-term observations become readily available, since having both temporal and frequency DM variations can break the degeneracy.  We will save the detailed question of a frequency-dependent DM for future investigation, as it is important for those who will correct high frequency data with DMs measured at low frequencies.  Although we offer no solution to the problem of disentangling profile evolution and dispersion measure, we can give greater credibility to measurements of dispersion measure changes, which are the more important quantities for timing experiments, and perhaps more interesting for studies of the ISM.

\subsubsection{Comparison of Measurement Uncertainties}
\label{TOAerr}

\begin{figure*}
\epsscale{1.0}
\plotone{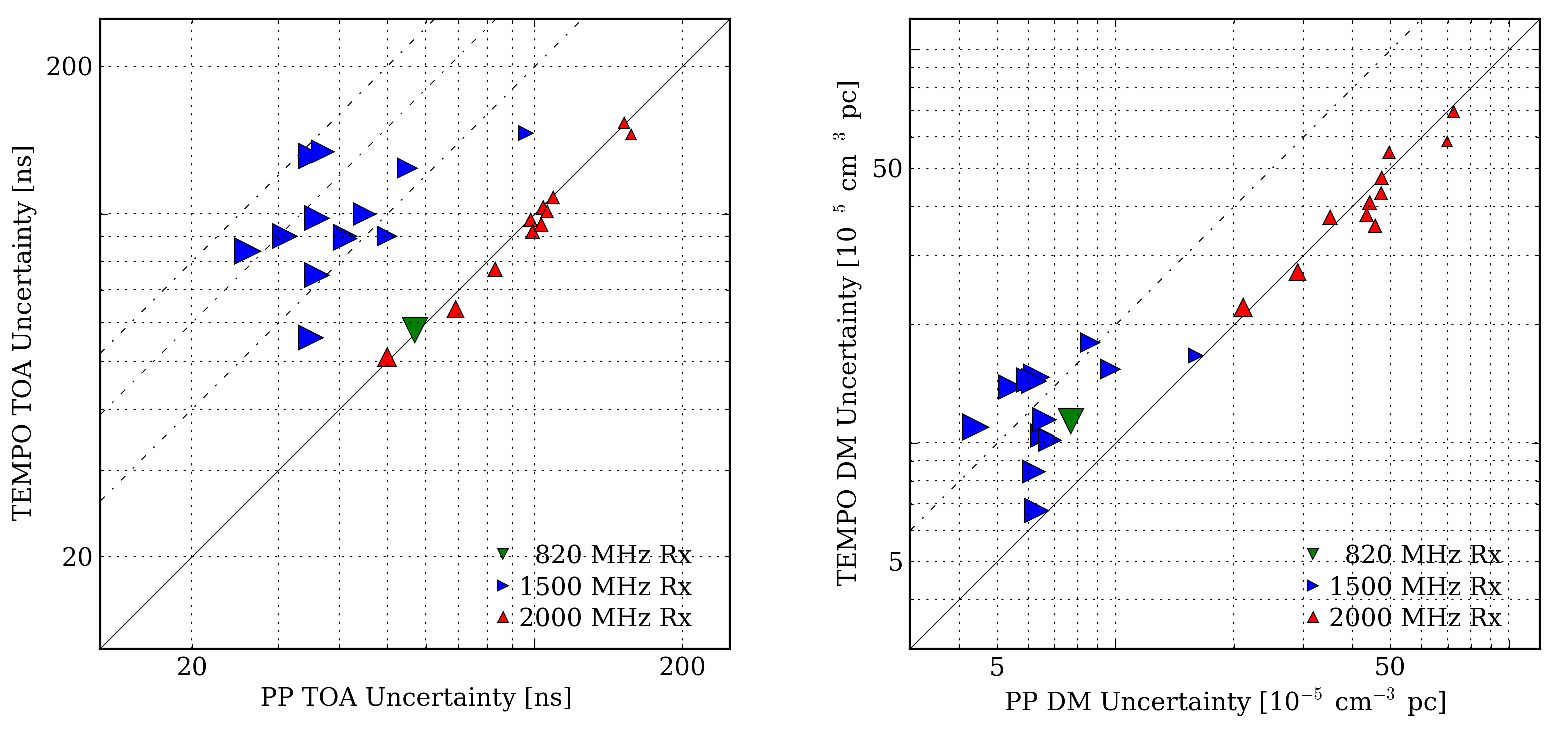}
\caption{A log-scale comparison of the TOA and DM uncertainties (left and right, respectively) from the M28A data.  The dash-dot lines indicate differences by factors of two (both panels), three, and four, and the area of each triangle is proportional to the data's SNR.  The largest improvements are seen in the 1500~MHz data, where profile evolution has been most mitigated.  The ``PP'' TOA uncertainties have been transformed to the same set of reference frequencies.  Two points in the left plot have the same values, so it appears as though there are only twelve 1500~MHz epochs.  The DM uncertainties have been scaled by their individual reduced $\chi^2$ values from the DM fit.}
\label{M28A_error_compare}
\end{figure*}

Figure~\ref{M28A_error_compare} shows a comparison of the uncertainties on the TOAs (left panel) and DMs (right panel).  The TOA uncertainties shown here are from frequency-averaged TOAs (one TOA per epoch, per band) that were obtained in a similar fashion as the multi-channel TOAs.  The DMs measured from the multi-channel TOAs were used to align the profiles before frequency-averaging them; this is one traditional way of accounting for significant DM changes, which would otherwise smear the average profile and systematically inflate the TOA error by an amount related to $\Delta$DM.  Each of these TOAs will reference some specific frequency and will be covariant with the DM.  In order to make a fair comparison, we have plotted the transformed ``PP'' TOA uncertainties to reference these frequencies; the zero-covariance uncertainties are smaller by $\lesssim 20\%$.\footnote{Additionally, these results are virtually unchanged if we remake the ``PP'' measurements on 64-channel data, which is consistent with what we present from our Monte Carlo tests in \S\ref{mc_results}.}

However, we did not use a weighted frequency-average, neglecting any SNR variation across the band that might originate from ISM effects or profile evolution; this is a second effect that can lessen the timing precision in the standard protocol.  Weighting the multi-channel TOAs (or the pulse profiles) to obtain a single frequency-averaged TOA reduced the ``TEMPO'' TOA uncertainties in Figure~\ref{M28A_error_compare} by factors between one and three, bringing all of them to within a factor of two of the ``PP'' uncertainties.

Finally, systematic trends from profile evolution (see Figure~\ref{M28A_freqevol_compare}) will enlarge the uncertainties.  We have scaled the DM measurement uncertainties by the reduced $\chi^2$ value of the fit to each epoch's multi-channel TOAs to better reflect the residual scatter.  The effect on the timing and DM precision from marginalizing M28A's profile evolution is unambiguous in the 1500~MHz data, which may explain why there is no segregation between the mean values of the 1500~MHz and 2000~MHz TOA uncertainties despite the former having significantly better SNRs.  The TOA uncertainties obtained by the new algorithm using a simple Gaussian model are smaller by up to a factor of four, with an average of about five-halves, and the DM uncertainties are smaller by up to a factor of two and a half, with an average of about three-halves.  The DM uncertainties are uniformly better at 1500~MHz because of the larger fractional bandwidth.  There is no improvement in either the TOA or DM uncertainties at 2000~MHz, since there is no significant profile evolution.

The ad hoc methods to mitigate effects arising from dispersion measure changes, frequency-dependent SNRs, and profile evolution in wideband data are all naturally accounted for by using the new algorithm, which we have seen to yield superior, or at least as good, measurement precisions.

\section{Monte Carlo Analyses}
\label{mcs}

\subsection{Description}

We completed a variety of Monte Carlo analyses to explore the accuracy to which the algorithm can determine parameter estimates, errors, and covariances in a number of regimes, which included varying the data resolution, the signal-to-noise ratio, scintillation patterns, and level of $\delta$DM.  Here, we show results from generating fake pulsar data by adding random, frequency-independent noise to the total-intensity model for M28A given in Table~\ref{model}.  The data emulate those from typical pulsar timing observations with GUPPI; we set the center frequency to 1500~MHz and the bandwidth to 800~MHz.  

We explored the performance of the algorithm in a variety of data resolutions, changing the number of phase bins in the profile from 128 to 2048, and the number of frequency channels from 8 to 512, both in powers of two.  When using an insufficient number of phase bins to resolve the profile, some harmonic power gets aliased into the estimate of the profile's noise level, which in turn suppresses the estimate of the parameter errors.  Since we expect these issues to be avoided in practice, and because our results seemed independent of the number of profile bins once the profile is resolved, we restrict the presentation of the Monte Carlo trials to those with profiles having $n_{bin} = 2048$.

For each sample in a given Monte Carlo trial, a random infinite-frequency phase was drawn uniformly from the interval [-0.5, 0.5) and injected into the model.  The injected DM value was the nominal ephemeris value plus a perturbation drawn uniformly from the log$_{10}$ interval of approximately [-5.0, -1.5], with equal probability given to the sign of the perturbation.  We chose this interval because it equally samples different scales of perturbations with a maximum that roughly corresponds to $\frac{\Delta\textrm{DM}}{P_s} \approx\ 100\ [10^{-4}$~cm$^{-3}$~pc ms$^{-1}$], and because we do not expect in most cases that a DM will be determined to better than $\sim$10$^{-5}$~cm$^{-3}$~pc. For simplicity, at a given SNR the RMS noise level remained constant as a function of frequency across the band, but in some of the tests we added random amplitude patterns ($a_n$) to mimic the effects of scintillation (not presented here, but see Figure~\ref{M28A_scint_demo} for an example).

\subsection{Results}
\label{mc_results}
\begin{figure}
\epsscale{1.0}
\plotone{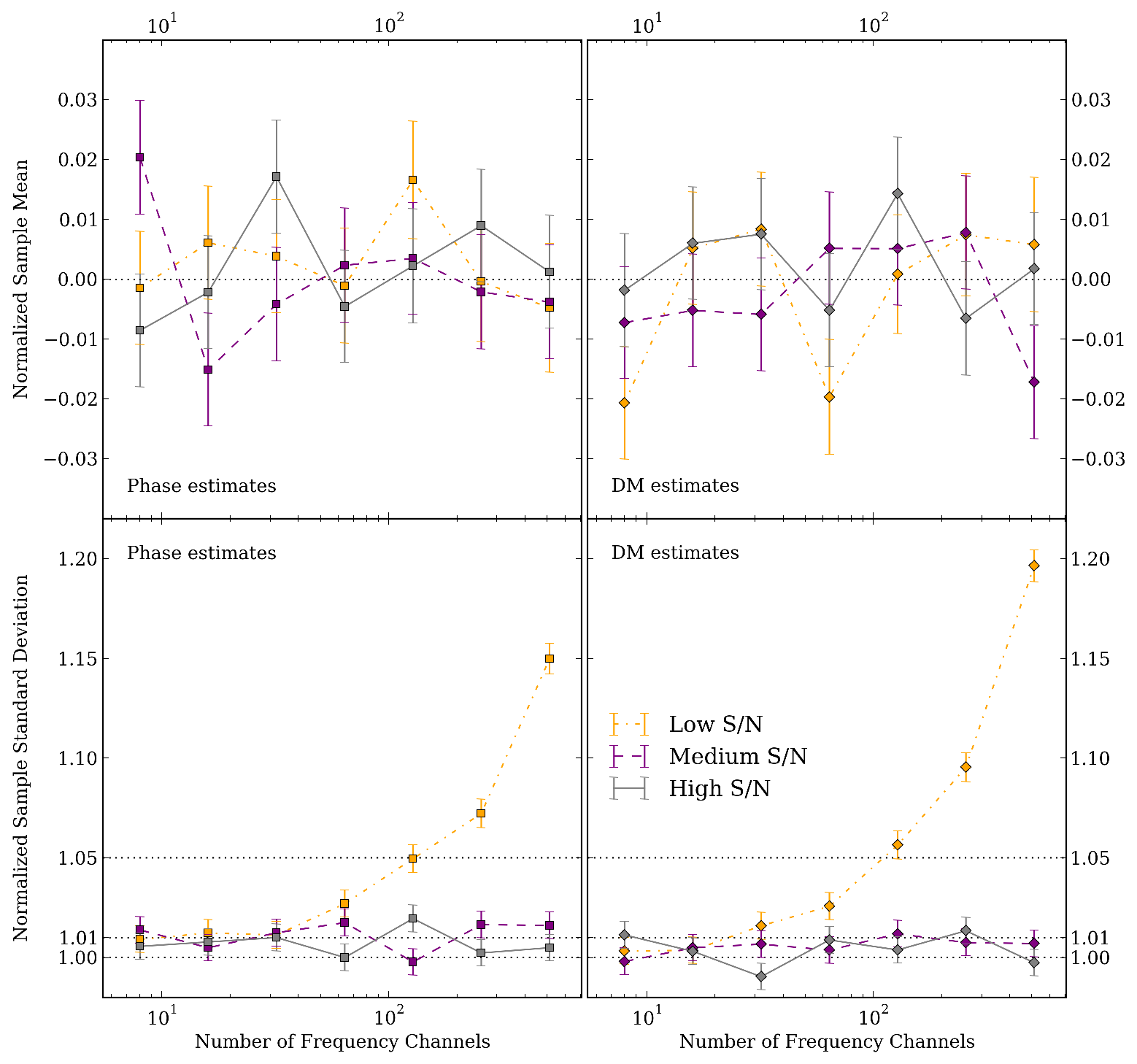}
\caption{Monte Carlo results for examining bias and error calculation as a function of $n_{chan}$ and SNR.  Each trial consisted of 11,400 samples.  The squares (left column) show results in each of three SNR regimes for the phase estimates, and the diamonds (right column) show the same for the DM estimates.  The two statistics shown are the normalized sample mean (top row) and standard deviation (bottom row).  The error bars are each one standard error of their respective statistic.  The dotted lines in the bottom row correspond to a 0\%, 1\%, and 5\% underestimation of the errors.  Additional Monte Carlo trials were performed for a wider range of SNRs, which fill the gap between the Medium and Low SNRs shown here, as well as perform even more poorly than the Low SNR trial.}
\label{MC_plot1}
\end{figure}

\begin{figure}
\epsscale{0.85}
\plotone{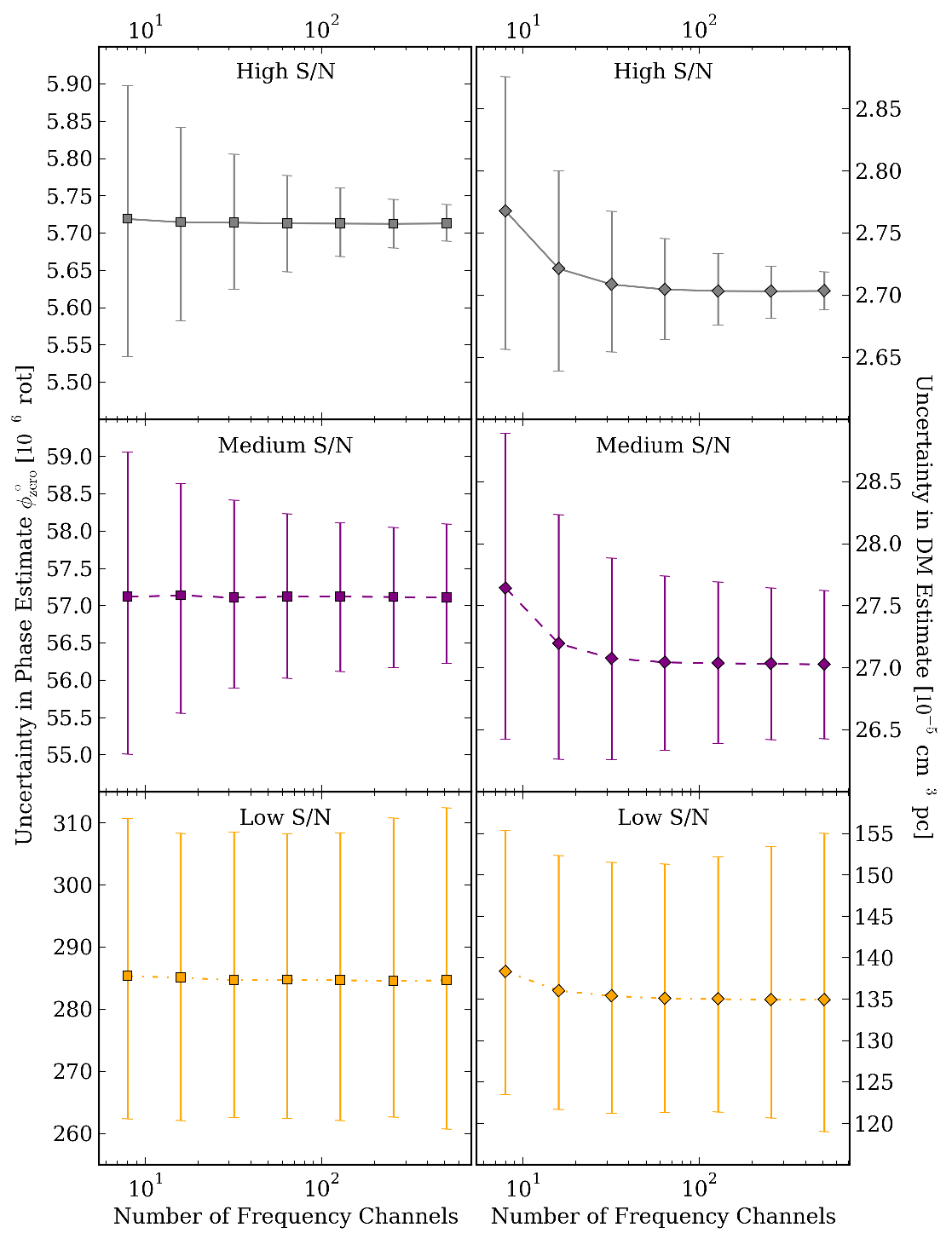}
\caption{The error distributions' dependency on $n_{chan}$ is plotted for the three SNR regimes. The plotted points show the distributions' median values and 95\% highest-density regions.  There is a slight skew in the error distribution for the Low SNR regime that becomes conspicuous at even lower SNRs.  Additional Very Low SNR Monte Carlo trials not shown here have severely skewed error distributions.}
\label{MC_plot2}
\end{figure}

\begin{figure}
\epsscale{1.0}
\plotone{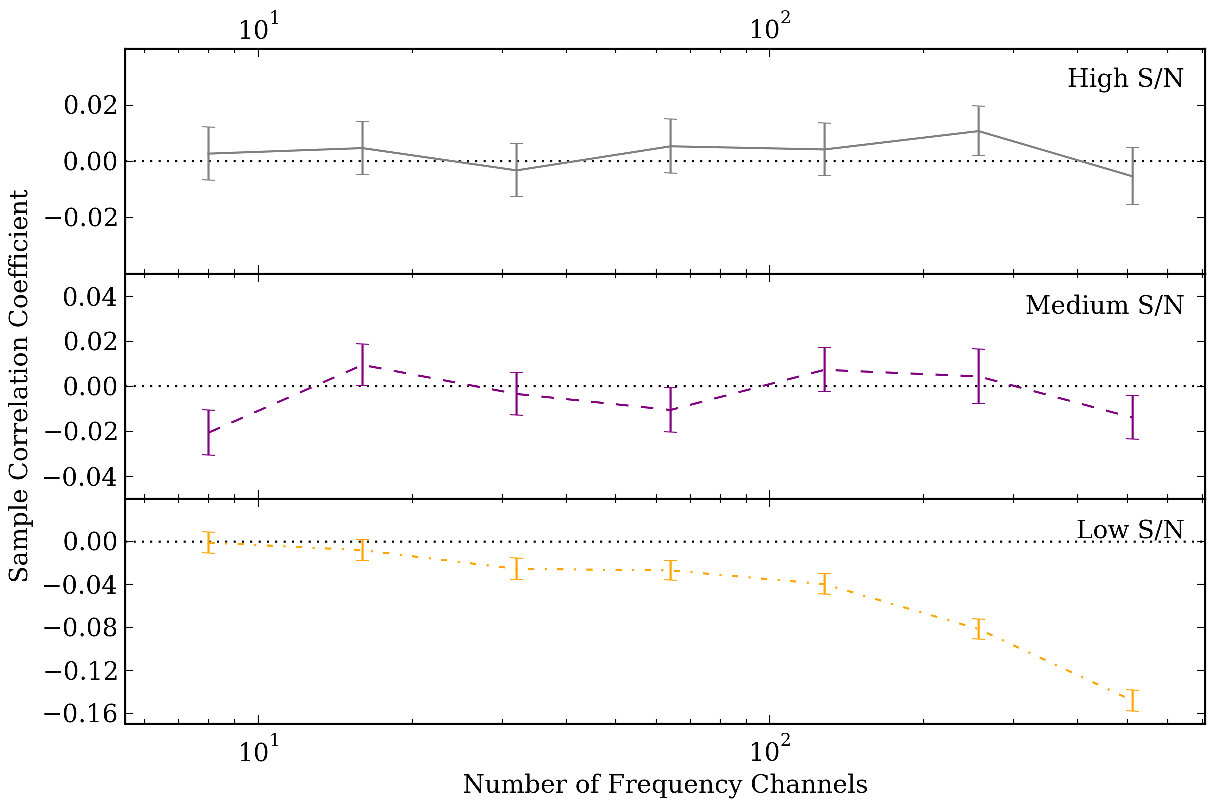}
\caption{The samples' normalized covariances are plotted to verify that $\nu_{zero}$ is the zero-covariance reference frequency.  The point estimates for the sample correlation coefficients and their errors were determined by a resampling analysis of each full Monte Carlo trial.  The trend prevalent in the Low SNR regime is due to an inaccurate determination of $\nu_{zero}$ (see text and Figure~\ref{MC_plot4}).}
\label{MC_plot3}
\end{figure}

Figures~\ref{MC_plot1}, \ref{MC_plot2}, and~\ref{MC_plot3}, show results from the Monte Carlo trials in three SNR regimes for the seven values of $n_{chan}$.  We used \texttt{PSRCHIVE} to measure the noise level in the data, and the three SNR levels in our trials presented here were set to be near the \texttt{PSRCHIVE} values of 20.0 (``Low''; yellow, dash-dot), 100.0 (``Medium''; purple, dashed), and 1000.0 (``High''; gray, solid)\footnote{The SNRs of the M28A data presented in the previous section varied between $\sim$100 and $\sim$700.}.

Figure~\ref{MC_plot1} shows two statistics returned from the trials for the phase estimates (left column, squares) and DM estimates (right column, diamonds) as a function of $n_{chan}$.  The top row shows the mean of the distribution of the values
\begin{equation}
\label{norm}
    \frac{\textrm{estimated value} - \textrm{injected value}}{\textrm{calculated error}},
\end{equation}
and the bottom row shows the standard deviation of this distribution.  If there are no systematic differences and if the errors are calculated accurately, then this distribution should be~$\sim\textrm{Normal}(0,1)$.  There is no obvious evidence of bias as a function of $n_{chan}$ or SNR, meaning the injected values are accurately recovered, within the error.  Even though all of the recovered normalized distributions were very well approximated by a Normal distribution (down to very small SNRs), one can see that the errors are underestimated when the channel-SNR becomes sufficiently low.  However, even in the Low SNR case with the largest number of channels, the errors are off by no more than 20\%.

Figure~\ref{MC_plot2} shows how the absolute errors change with $n_{chan}$.  We have separated the trials for clarity; each point represents the median of the error distribution, contained within the 95\% highest-density region.  The uncertainty scales linearly with the SNR for both parameters.  There is no obvious dependence on the average TOA error with $n_{chan}$, but there is some increase in the DM error by a few percent as the number of channels becomes small, independent of the SNR.  This is expected after considering Equation~\ref{covar2} because the effective frequency range of the data (the difference of the center frequencies in the highest and lowest channels) decreases with the number of channels as $n_{chan} ^{\ \ -1}$; that is, the ``lever arm'' for the DM measurement lessens, giving a greater measurement uncertainty.

The non-Gaussianity of the error distributions in Figure~\ref{MC_plot2} is somewhat noticeable in the Low SNR regime towards higher $n_{chan}$, but it is manifested in the trend seen in the lower half of Figure~\ref{MC_plot1}.  The skewness towards small uncertainties becomes very obvious at lower SNRs (not shown here).  The underestimation of TOA errors at low SNRs (particularly for profiles with large duty-cycles) has been documented before \citep{Hotan05}.

Lastly, there is a dependency of each error distribution's variance on $n_{chan}$.  Beyond some low channel-SNR, the variance appears constant, which has been verified for the Middle SNR case.  The variance of the error distribution is not a particularly interesting quantity, so we will refrain from additional discussion, making only a note that it seems to affect the value of the errors at the level of a few percent.  After replicating this series of Monte Carlo trials with a similar Gaussian model that has no profile evolution (i.e. the same components with constant positions, widths, and amplitudes), we find almost identical results, except that the absolute errors uniformly decreased by $\sim$3\%.  From this we conclude that the effects from marginalizing over profile evolution in each channel for this model were minimal.

It is important to remember that all phase and error estimates shown here are referenced to $\nu_{zero}$; if a different reference frequency were used, the results in Figures~\ref{MC_plot1} and~\ref{MC_plot2} would look different because of non-zero covariance between $\phi^{\circ}_{ref}$ and DM.  To verify if the calculated $\nu_{zero}$ is, in fact, the zero-covariance dedispersion reference frequency, we show the sample correlation coefficient (the sample covariance normalized by the sample standard deviations) in Figure~\ref{MC_plot3} for the same three sets of Monte Carlo trials.  It is obvious that there is deviation from zero covariance in the Low SNR regime.

\begin{figure}
\epsscale{0.8}
\plotone{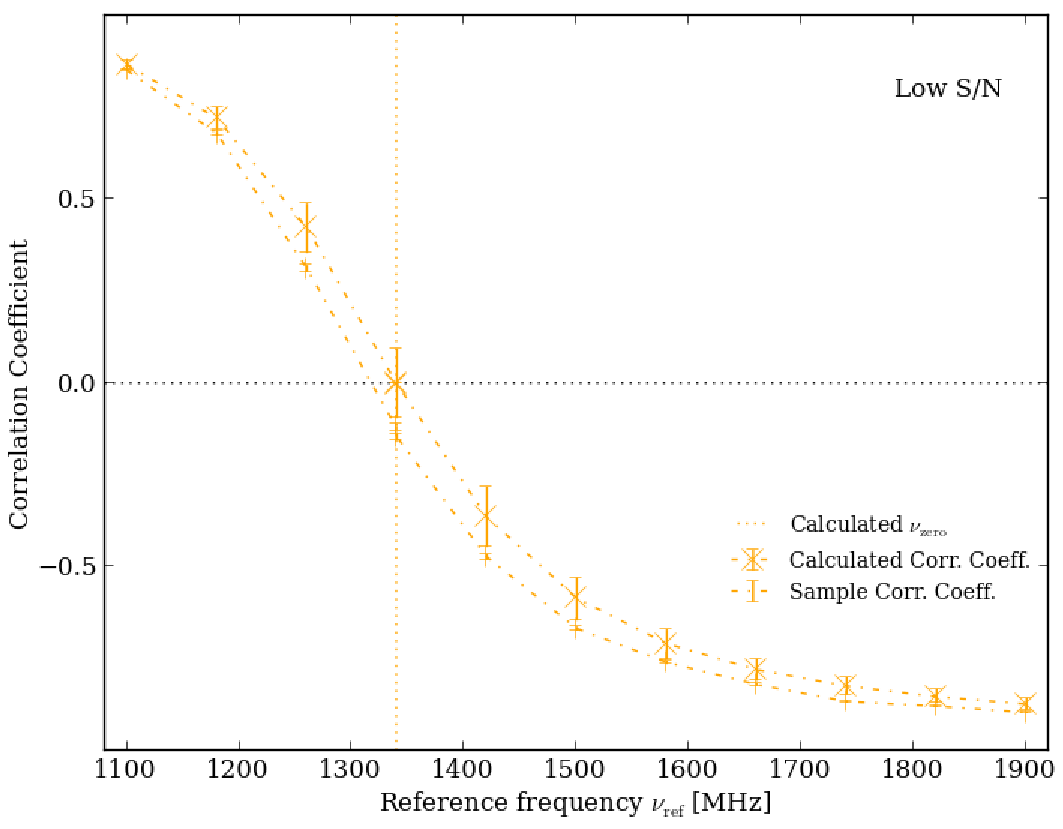}
\caption{The Low SNR correlation coefficient for Monte Carlo trials as a function of $\nu_{ref}$ for $n_{chan} = 512$.  The vertical dotted line shows the calculated value for $\nu_{zero}$ (the average value is plotted, but there is effectively zero dispersion).  The interpolated value for the ``true'' $\nu_{zero}$ in the sample differs by $\sim$30~MHz.  Note that the sample correlation coefficient at $\nu_{zero}$ here agrees with that from Figure~\ref{MC_plot3} for $n_{chan} = 512$.  The equivalent curves for the higher SNR trials overlap almost exactly.  The point estimates of the sample correlation coefficients and their errors were determined in the same way as Figure~\ref{MC_plot3}.}
\label{MC_plot4}
\end{figure}

One underlying issue that can explain this feature is that our analytic formulation for $\nu_{zero}$ given in Equation~\ref{nuzero} will not be precise for all data resolutions of arbitrary SNR\footnote{Another way to say this is that the likelihood function associated with Equation~\ref{chi2} becomes non-Gaussian in the low SNR limit.}.  This is verified for the low SNR case in Figure~\ref{MC_plot4}, which shows the discrepancy between our calculated covariances and the covariances measured in additional Monte Carlo samples.  The Monte Carlo trials are the same as before, but are now fixed with $n_{chan} = 512$, while varying $\nu_{ref}$.  The vertical dotted line shows the calculated $\nu_{zero}$ for this SNR, which is significantly offset ($\sim$30~MHz) from the interpolated zero-crossing of the sample correlation coefficient curve.  That the slopes of the functions in Figure~\ref{MC_plot4} are steepest near the zero-crossing implies that the determination of the parameter uncertainties is sensitive to the determination of $\nu_{zero}$.

\section{Conclusions}
\label{conc}
In this paper, we have presented a simple method for measuring TOAs in folded pulsar data by using a frequency-dependent model of the pulse profile.  This algorithm is a straightforward, yet novel extension of FFTFIT from~\citet{Taylor92}, but it has some advantages over more standard techniques.  These include:
\begin{itemize}
\item{Simultaneous measurement of a phase (TOA) and dispersion measure (DM) across a channelized bandwidth, allowing for easy DM tracking.}
\item{{\it In situ} accommodation for profile evolution, both intrinsic and extrinsic (eg. scattering), by the use of an arbitrary phase-frequency model (a ``portrait'') of the pulsar's signal.}
\item{Mitigation of scintillation effects by automatic weighting of frequency channels, which appropriately uses the information contained in the data.}
\item{Simplification of the timing procedure by eliminating (where appropriate) the use of phase offsets (i.e. JUMPs), multi-channel TOAs, and ad hoc methods for describing profile evolution, etc.}
\end{itemize}
Any arbitrary model can be used, but the choice of model will affect the measured values.  We have made our code publicly available online.

The demonstration of a simple Gaussian modeling scheme to make these measurements in a 3-year, wideband dataset of the millisecond pulsar M28A shows that we are able to obtain reliable measurements of the dispersion measure, as well as improved TOA and DM precisions by up to a factor of four for the former and two for the latter.  The biggest improvements in the parameter precisions and in mitigating profile evolution were seen in the high signal-to-noise 1500~MHz data, which was our best ``wideband data'' for demonstration purposes in the sense that it has the largest fractional bandwidth (and therefore the most obvious effects from interstellar dispersion, scattering, and profile evolution).  We note that M28A was chosen for our demonstration precisely because it would show obvious improvements from using our new technique, and so its results may be atypical.  Similar improvements in other pulsars will depend on the mitigation of intrinsic and extrinsic profile evolution.  It became clear in our comparisons that there is a necessity for quantitative model selection based on more robust two-dimensional portrait modeling, which potentially can lead to the detection of a frequency-dependent DM, or other interesting signals.

We probed three typical SNR regimes with Monte Carlo tests and found that the algorithm performs well.  Except for the lowest SNR cases, the results from our Monte Carlo analyses has led us to the conclusion that a large number of frequency channels is appropriate for applying this technique.  A larger number of channels will provide the highest precision DM measurements and avoids averaging over profile evolution.  The proper incorporation of discrete DM measurements with their own heteroscedastic errors (besides the TOAs') into the determination of a timing model (eg. by using \texttt{tempo}) is not trivial, but a Bayesian approach has been investigated in~\citet{Lentati13}.  Relatedly, measuring DMs from non-simultaneous but temporally proximate multi-frequency data can be an intermediate improvement until larger bandwidths become readily available.  This is another avenue of future development, although it comes with the drawback of having correlated TOAs.

One important caveat in these Monte Carlo tests is that the model fitted to the simulated data was the true model from which the data were generated.  In practice, a Gaussian-component model fitted to real data will not match perfectly (leaving behind non-Gaussian residuals in all but the simplest or low SNR cases) and there will be a much stronger dependence of the measured DM and TOA on the number of channels and the amount of profile evolution.  This is another area requiring further testing, but it also suggests to err on the side of more frequency channels.

Although general, the algorithm will be most useful when applied to MSPs because of their sensitivities to small dispersion measure changes, as highlighted by Equation~\ref{DMsmear}, and because of the need to correct for their profile evolution in wideband data to obtain the highest possible timing precisions.  For these reasons, we believe this algorithm will provide a natural TOA and DM measurement procedure for campaigns of MSP monitoring, like that of NANOGrav or other PTA experiments.  To determine how much is gained in timing precision (and, perhaps, sensitivity to gravitational waves), a direct application of our timing method to the wideband data from a subset of NANOGrav MSPs will be presented in a future paper.

\acknowledgments
The authors would like to thank A. Bilous for extensive help with the M28A data and comments on the manuscript, as well as P. Freire, J. Hessels, R. Lynch, and I. Stairs, who were additionally responsible for proposing the M28A observations and obtaining the data.  Gratitude is also expressed to the NANOGrav collaboration.  TTP is supported in part by a National Science Foundation PIRE Grant (0968296) through NANOGrav.

%K\"{o}sz\"{o}n\"{o}m sz\'{e}pen.

Facility: \facility{GBT (GUPPI)}

\appendix

\section{Covariance and Error Estimates of Fitted Parameters}
\label{apndx}

We will denote our least-squares estimate of the parameters $\theta = \{\phi^\circ_{ref}, \textrm{DM}, a_n\}$ by $\hat{\theta}$.  Equation~\ref{chi2_2} has a minimum at $\hat{\theta} = \{\hat{\phi}^\circ_{ref}, \hat{\textrm{DM}}, \hat{a}_n\}$, where $\hat{a}_n = \frac{\hat{C}_{dp,n}}{S_{p,n}} = \frac{C_{dp,n}(\hat{\phi}^\circ_{ref}, \hat{\textrm{DM}})}{S_{p,n}}$ (cf. Equation~\ref{an}).  We can approximate the $(n_{chan}\!+\!2) \times (n_{chan}\!+\!2)$ covariance matrix of the parameters by the inverse of the curvature matrix $\boldsymbol{\kappa}$, which can be derived from a Taylor expansion of the $\chi^2$ function near its minimum.  The entries of the curvature matrix at the minimum point are given by
\begin{equation}
\label{kappa}
    \kappa_{kl} = \frac{1}{2}\frac{\partial^2\chi^2(\theta)}{\partial\theta_k\partial\theta_l}\bigg|_{\hat{\theta}},
\end{equation}
which is one-half of the Hessian.  However, because we are primarily interested in the errors and covariances of $\hat{\phi}^\circ_{ref}$ and $\hat{\textrm{DM}}$, we will only calculate here the terms of the $2 \times 2$ Hessian for an arbitrary point of the function $\chi^{2}(\phi^\circ_{ref},\textrm{DM})$ given in Equation~\ref{chi2}.  Inverting this matrix to arrive at the $2 \times 2$ covariance matrix of interest is trivial, particularly because there is a reference frequency $\nu_{zero}$ that gives zero covariance between $\hat{\phi}^\circ_{ref}$ and $\hat{\textrm{DM}}$\footnote{To clarify, the covariances with the $\hat{a}_n$ estimates are already included in the $2 \times 2$ covariance matrix; there is only zero covariance at $\nu_{zero}$ between the fitted phase and DM.}.  One can arrive at the same results for the corresponding entries of the ``full'' covariance matrix by inverting the matrix in Equation~\ref{kappa} and inserting the values $a_n = \frac{C_{dp,n}}{S_{p,n}}$.  The three unique second-derivatives of Equation~\ref{chi2} are 
\begin{subequations}
\label{covar}
\begin{equation}
\label{covar1}
    \frac{\partial^2\chi^2}{\partial\phi^{\circ2}_{ref}} = -2\sum_nw_n,
\end{equation}
\begin{equation}
\label{covar2}
    \frac{\partial^2\chi^2}{\partial\textrm{DM}^2} = -2\sum_nw_n\Big[\frac{K}{P_s}(\nu^{-2}_n - \nu^{-2}_{ref})\Big]^2,
\end{equation}
and
\begin{equation}
\label{covar3}
    \frac{\partial^2\chi^2}{\partial\phi^\circ_{ref}\partial\textrm{DM}} = -2\sum_nw_n\Big[\frac{K}{P_s}(\nu^{-2}_n - \nu^{-2}_{ref})\Big],
\end{equation}
\end{subequations}
where
\begin{equation}
\label{wn}
    w_n \equiv S^{-1}_{p,n}\times(C^{\prime2}_{dp,n} + C^{\prime\prime}_{dp,n}C_{dp,n}),
\end{equation}
and the derivatives of $C_{dp,n}$ are with respect to $\phi_n$.  Requiring that the cross-term (Equation~\ref{covar3}) be equal to zero leads us to the zero-covariance dedispersion reference frequency $\nu_{zero}$,
\begin{equation}
\label{nuzero}
    \nu_{zero} = \sqrt{\frac{\sum_nw_n}{\sum_nw_n\nu^{-2}_n}}.
\end{equation}
Because Equation~\ref{Cdpn} is a function of $\phi_n$, we can transform the least-squares estimate $\hat{\phi}^\circ_{ref}$ to an estimate that has zero covariance with $\hat{\textrm{DM}}$ and ensure we are still at the minimum point,
\begin{equation}
\label{phizero}
    \hat{\phi}^\circ_{zero} = \hat{\phi}^\circ_{ref} + \Big[\frac{K\times\hat{\textrm{DM}}}{P_s}\Big(\hat{\nu}^{-2}_{zero} - \nu^{-2}_{ref}\Big)\Big].
\end{equation}
Under this transformation of the $\phi^\circ_{ref}$ coordinate, the Hessian is diagonal and the variances of the estimates $\hat{\phi}^\circ_{zero}$ and $\hat{\textrm{DM}}$ are simply twice the inverse of Equations \ref{covar1} and \ref{covar2}, respectively.  A derivation starting with the full $(n_{chan}\!+\!2) \times (n_{chan}\!+\!2)$ Hessian confirms that these errors incorporate the covariances with the $a_n$.  Furthermore, the Monte Carlo results from~\S\ref{mcs} and Figure~\ref{MC_plot3} show that we have accurately been able to calculate covariances down to low SNR levels.  The default output of our \texttt{python} code references the phase estimates and TOAs to $\nu_{zero}$, which is our recommendation for any similar implementation.

\bibliography{apj-jour,PDR13_arxiv}

\end{document}